\documentclass[12pt, a4paper]{article}
\usepackage{graphicx}
\usepackage{amssymb}
\usepackage{amsmath}
\usepackage{bm}
\usepackage{color}
\usepackage{theorem}
\usepackage{subcaption}
\usepackage{listings}
\usepackage{colortbl}
\usepackage{tabularx}
\usepackage{longtable}
\usepackage[utf8]{inputenc}
\usepackage[T1]{fontenc}
\usepackage{lmodern}
\usepackage[top=30truemm,bottom=30truemm,left=25truemm,right=25truemm]{geometry}
\usepackage[sort&compress,numbers, merge]{natbib}
\usepackage{multirow}

\definecolor{Orange}{cmyk}{0,0.61,0.87,0}
\definecolor{JungleGreen}{cmyk}{0.99,0,0.52,0}
\definecolor{OliveGreen}{cmyk}{0.64,0,0.95,0.40}
\definecolor{Brown}{cmyk}{0,0.81,1,0.60}
\definecolor{RoyalBlue}{cmyk}{0.71,0.53,0,0.12}
\definecolor{Gray}{cmyk}{0,0,0,0.40}
\definecolor{LightPink}{cmyk}{0.0,0.25,0,0}
\definecolor{LLightPink}{cmyk}{0.0,0.10,0,0}
\definecolor{LightBlue}{cmyk}{0.25,0,0,0}
\definecolor{LightGray}{cmyk}{0,0,0,0.2}

\renewcommand{\arraystretch}{1.3}
\allowdisplaybreaks[1]

\usepackage[colorlinks=true, linkcolor=OliveGreen, citecolor=RoyalBlue,
urlcolor=RoyalBlue]{hyperref}

\renewcommand{\thefootnote}{\fnsymbol{footnote}}

\begin{document}

\begin{titlepage}

  \begin{flushright}
    {\tt
 ~~
    }
\end{flushright}

\vskip 1.35cm
\begin{center}

{\LARGE
{\bf
Capture of Electroweak Multiplet \\[5pt] Dark Matter in Neutron Stars
}
}

\vskip 1.5cm

Motoko Fujiwara$^{a,b}$\footnote{
  E-mail address: \href{mailto:motoko@hep-th.phys.s.u-tokyo.ac.jp}{\tt motoko@hep-th.phys.s.u-tokyo.ac.jp}},
Koichi~Hamaguchi$^{b,c}$\footnote{
  E-mail address: \href{mailto:hama@hep-th.phys.s.u-tokyo.ac.jp}{\tt hama@hep-th.phys.s.u-tokyo.ac.jp}}, 
Natsumi Nagata$^b$\footnote{
E-mail address: \href{mailto:natsumi@hep-th.phys.s.u-tokyo.ac.jp}{\tt natsumi@hep-th.phys.s.u-tokyo.ac.jp}},
and
Jiaming~Zheng$^d$\footnote{
  E-mail address: \href{mailto:zhengjm3@sjtu.edu.cn}{\tt zhengjm3@sjtu.edu.cn}}

\vskip 0.8cm

{\it $^a$Department of Physics, Nagoya University, Chikusa-ku, Nagoya, 464--8602 Japan}\\[2pt]
{\it $^b$Department of Physics, University of Tokyo, Bunkyo-ku, Tokyo
 113--0033, Japan} \\[2pt]
 {\it $^c$Kavli IPMU (WPI), University of Tokyo, Kashiwa, Chiba
  277--8583, Japan} \\[2pt]
 {\it $^d$School of Physics and Astronomy,
 Shanghai Jiao Tong University, Shanghai 200240, China}

\date{\today}

\vskip 1.5cm

\begin{abstract}

    If dark matter has a sizable scattering cross section with nucleons, it can efficiently be captured by a neutron star. Its energy is then transferred to the neutron star as heat through the scattering and annihilation inside the star. This heating effect may be detectable via dedicated temperature observations of nearby old pulsars, providing an alternative method for dark matter searches. In this paper, we show that for electroweak multiplet dark matter this search strategy can probe the parameter region which is out of reach of future dark matter direct detection experiments. To see this systematically, we classify such dark matter candidates in terms of their electroweak charges and investigate the effect of ultraviolet physics by means of higher-dimensional effective operators. We then show that if the effect of ultraviolet physics is sizable, the dark matter-nucleon elastic scattering cross section becomes sufficiently large, whilst if it is suppressed, then the mass splittings among the components of the dark matter multiplet get small enough so that the inelastic scattering processes are operative. In any case, the electroweak multiplet dark matter particles are efficiently captured in neutron stars, making the search strategy with the temperature observation of old neutron stars promising. 

\end{abstract}

\end{center}
\end{titlepage}

\renewcommand{\thefootnote}{\arabic{footnote}}
\setcounter{footnote}{0}

\section{Introduction}

An electrically neutral, stable particle that has a mass of ${\cal O}(100)$~GeV--${\cal O}(1)$~TeV and couples to the Standard Model (SM) particles with a strength similar to the electroweak interactions can be a good dark matter (DM) candidate, as its thermal relic abundance can naturally explain the observed DM density, $\Omega_{\rm DM} h^2 \simeq 0.12$~\cite{Planck:2018vyg}. A simple realization of such a particle is provided by an SU(2)$_L$ multiplet with a mass in that range, whose neutral component is the DM candidate. This DM interacts with the SM particles via the weak interaction, and its stability is assured by a discrete symmetry or emerges accidentally due to the large dimension of the representation~\cite{Cirelli:2005uq, Cirelli:2007xd, Cirelli:2009uv}. We refer to this class of DM candidates as \textit{electroweak multiplet dark matter}. Electroweak multiplet dark matter often appears in new physics models. Well-known examples are higgsino and wino in supersymmetric models, which are SU(2)$_L$ doublet and triplet fermions, respectively. Grand unified theories based on an SO(10) or E$_6$ gauge group may also contain this class of DM, both scalars and fermions, with its stability explained by a remnant discrete symmetry of the unified gauge group~\cite{Kadastik:2009dj, Kadastik:2009cu, Frigerio:2009wf, Hambye:2010zb, Mambrini:2013iaa, Mambrini:2015vna, Nagata:2015dma, Arbelaez:2015ila, Boucenna:2015sdg, Evans:2015cqq, Mambrini:2016dca, Parida:2016hln, Nagata:2016knk, Schwichtenberg:2017xhv, Ma:2018uss, Ferrari:2018rey, Abe:2021byq, Cho:2021yue}.  A spin-1 candidate for the electroweak multiplet DM can appear in a theory with an extended gauge group~\cite{Abe:2020mph}.

The electroweak multiplet DM has several characteristic features, which have important implications for its phenomenology. In particular, if the DM is a fermion, its properties are determined robustly as there is no renormalizable coupling to the Higgs field; we focus on this case in this paper just for simplicity, but an extension of our study to the scalar and vector DM cases is straightforward. For fermionic electroweak multiplet DM, the DM phenomenology is basically determined by the electroweak gauge interactions. It is found that all of the components in the DM multiplet are degenerate in mass at tree level, and the electroweak radiative corrections generate mass splittings  of ${\cal O}(100)$~MeV among the components, making the neutral component the lightest. Moreover, the effective annihilation cross section of the DM particles is highly enhanced by coannihilation~\cite{Griest:1990kh} with other components in the DM multiplet as well as by the Sommerfeld enhancement effect~\cite{Hisano:2003ec, Hisano:2004ds}. As a result, the thermal relic abundance agrees to the observed DM density for a relatively large DM mass, ${\cal O}$(1--10)~TeV. Another feature of this class of DM candidates is that, as long as renormalizable couplings are concerned, their scattering with nucleons is only induced at loop level~\cite{Hisano:2004pv, Hisano:2010fy, Hisano:2010ct, Hisano:2011cs, Hill:2013hoa, Hill:2014yxa, Hisano:2015rsa}, and thus its cross section is rather small. For instance, the latest calculation~\cite{Hisano:2015rsa} shows that the DM-nucleon spin-independent (SI) scattering cross section is $\simeq 8 \times 10^{-50}~\mathrm{cm}^2$, $2 \times 10^{-47}~\mathrm{cm}^2$, and $2 \times 10^{-46}~\mathrm{cm}^2$, for the cases of doublet, triplet, and quintuplet, respectively; for the triplet and quintuplet cases, a multi-ton scale detector is required to test these predictions in DM direct detection experiments, while for the doublet case, the resultant scattering cross section is below the neutrino floor~\cite{Billard:2021uyg}---where it becomes challenging to detect the DM signal due to the neutrino background. 

Although the electroweak multiplet DM is elusive on the earth, it can leave a significant imprint on neutron stars (NSs). Because of the large gravitational field generated by a NS near its surface, a DM particle obtains a large velocity ($v \gtrsim 0.5 c$) when it collides with a NS. It can exchange a momentum of $\gtrsim {\cal O}(100)$~MeV with the NS matter in the collision, with which inelastic scattering into another component can occur via the electroweak gauge boson exchange at tree level, if the transferred energy is larger than the mass difference between this component and the DM. The cross section of this inelastic scattering process is large enough for the DM to be captured by the NS efficiently. The captured DM particles eventually transfer their energy to the NS as heat through the scattering and annihilation. This heating effect~\cite{Kouvaris:2007ay, Bertone:2007ae, Kouvaris:2010vv, deLavallaz:2010wp, Baryakhtar:2017dbj} may be detectable through dedicated NS observations in the foreseeable future. Nearby cold NSs are likely to be among the continuously expanding set of new pulsars discovered by radio telescopes such as the Five-hundred-meter Aperture Spherical radio Telescope\,(FAST)\,\cite{Han:2021ekd}. Their temperature may be further measured by future observations with infrared telescopes such as the recently launched James Webb Space Telescope (JWST)~\cite{Gardner:2006ky}. For recent related studies, see Refs.~\cite{Bramante:2017xlb, Raj:2017wrv, Chen:2018ohx, Bell:2018pkk, Garani:2018kkd, Camargo:2019wou, Bell:2019pyc, Hamaguchi:2019oev, Garani:2019fpa, Acevedo:2019agu, Joglekar:2019vzy, Keung:2020teb, Yanagi:2020yvg, Joglekar:2020liw, Bell:2020jou, Ilie:2020vec, Bell:2020lmm, Bell:2020obw, Maity:2021fxw, Anzuini:2021lnv, Zeng:2021moz, Bramante:2021dyx, Tinyakov:2021lnt, Hamaguchi:2022wpz}.

It is, however, non-trivial if the inelastic scattering of the electroweak multiplet DM can actually occur since both the energy transfer in the scattering process and the mass difference between the DM and another component are expected to be ${\cal O}(100)$~MeV. In addition, if there are some extra heavy particles that couple to the DM multiplet, then the mass difference tends to be increased, which may suppress the inelastic scattering process. On the other hand, even if the inelastic scattering cannot occur, the DM can still be captured by NSs if the elastic scattering cross section is large enough, which might be the case if the extra particles induce DM-nucleon interactions. In this paper, we study this issue systematically based on the method of effective field theories~\cite{Hisano:2014kua, Nagata:2014wma, Nagata:2014aoa, Dedes:2016odh, Fukuda:2017jmk, Kuramoto:2019yvj, Geytenbeek:2020rxa, Belyaev:2022qnf}. We classify the electroweak DM candidates in terms of their $\mathrm{SU}(2)_L \times \mathrm{U}(1)_Y$ quantum numbers and take into account the effect of ultraviolet (UV) physics on the mass difference using higher-dimensional effective operators suppressed by a heavy mass scale, $\Lambda$. The condition for the occurrence of inelastic scattering is then obtained as a lower limit on the scale $\Lambda$. We also find complementary roles between elastic and inelastic scatterings for the capture of the electroweak multiplet DM in NSs; for a lower $\Lambda$, the mass splittings among the components in the DM multiplet get larger, suppressing inelastic scattering, but in the meantime the DM-nucleon elastic scattering cross section gets larger, which maintains the DM capture rate large enough.

The layout of this paper is as follows. We first discuss the DM heating effect on  NSs in Sec.~\ref{sec:heating}. In Sec.~\ref{eq:EMDM}, we review relevant features of the electroweak multiplet DM. In Sec.~\ref{sec:scattering}, we study the DM-nucleon scattering cross section for this class of DM candidates and discuss the sensitivities of the NS temperature observation and DM direct detection experiments. Finally, we summarize our results in Sec.~\ref{sec:conclusion}.

\section{Neutron Star Heating with Dark Matter}
\label{sec:heating}

DM particles near a NS are attracted by the gravitational field generated by the NS. They collide with the star if its impact parameter is smaller than~\cite{Goldman:1989nd}
\begin{equation}
  b_{\rm max} = R_{\rm NS} \cdot \frac{v_{\rm esc}}{v_{\rm DM}} \cdot e^{- \Phi (R_{\rm NS})} ~,
\end{equation}
where $R_{\rm NS}$ is the NS radius, $v_{\rm esc} = (2G M_{\rm NS}/R_{\rm NS} )^{1/2}$ is the escape velocity, $G$ is the gravitational constant, $M_{\rm NS}$ is the NS mass, $v_{\rm DM}$ is the DM velocity distant from the NS, and 
\begin{equation}
  e^{- \Phi (R_{\rm NS})} = \biggl(1 - \frac{2 G M_{\rm NS}}{R_{\rm NS}}\biggr)^{-1/2}~. 
\end{equation}
Under the assumption that the DM velocity follows the Maxwell-Boltzmann distribution with the mean squared velocity $\bar{v}^2_{\rm DM}$, the rate of the DM particles colliding into the NS is evaluated as~\cite{Kouvaris:2007ay} 
\begin{equation}
  \frac{dN}{dt} = \sqrt{\frac{6}{\pi}}  \frac{ \pi b^2_{\rm max} \bar{v}_{\rm DM}  \rho_{\rm DM}}{m_{\rm DM}} ~,
  \label{eq:dndt}
\end{equation} 
where $\rho_{\rm DM}$ and $m_{\rm DM}$ are the local energy density and mass of DM, respectively.

As it turns out, a DM particle with a mass $\lesssim 1$~PeV requires only one scattering to be captured by the NS, since the energy transfer in the scattering process tends to be much larger than the initial kinetic energy of the DM. To see this, we note that the energy deposit of DM in each scattering is evaluated as 
\begin{equation}
  \Delta E = \frac{m_N m_{\rm DM}^2 \gamma^2_{\rm esc} v_{\rm esc}^2}{m_N^2 + m_{\rm DM}^2 + 2 \gamma_{\rm esc} m_{\rm DM} m_N} (1 - \cos \theta_{\rm cm}) ~,
  \label{eq:deltae}
\end{equation}
where $m_N$ ($N=p,n$) denotes the nucleon mass, $\theta_{\rm cm}$ is the scattering angle in the center-of-mass frame, and $\gamma_{\rm esc} \equiv (1-v_{\rm esc}^2)^{-1/2}$. For a typical NS, the escape velocity $v_{\rm esc}$ is 
\begin{equation}
  v_{\rm esc} \simeq 2 \times 10^{8} \times \biggl(\frac{M_{\rm NS}}{1.4 M_{\odot}}\biggr)^{1/2} \biggl(\frac{R_{\rm NS}}{10~{\rm km}}\biggr)^{-1/2} ~{\rm m}/{\rm s} ~,
\end{equation}
\textit{i.e.}, as large as the speed of light. We then find that for $m_{\rm DM} \gg m_N$, the coefficient in front of the parenthesis in Eq.~\eqref{eq:deltae} is $\simeq m_N \gamma^2_{\rm esc} v_{\rm esc}^2$, which is much larger than the initial kinetic energy of DM, $m_{\rm DM} v_{\rm DM}^2/2$, for $m_{\rm DM} \lesssim 1$~PeV. 

The incoming DM particles experience at least one scattering in the NS if the mean free path, $(\sigma^{(N)} n_N)^{-1}$, is smaller than the size of the NS, $\sim R_{\rm NS}$, where $\sigma^{(N)}$ is the DM-nucleon scattering cross section and $n_N$ is the number density of nucleons. By approximating $n_N \sim M_{\rm NS}/(m_N R_{\rm NS}^3)$, we obtain a lower limit on $\sigma^{(N)}$ from the above condition as $\sigma^{(N)} \gtrsim m_N R^2_{\rm NS}/M_{\rm NS}$. 
The geometric cross section of the NS is often used to estimate the lower limit of the inequality, which defines the threshold cross section as\footnote{With the Akmal-Pandharipande-Ravenhall (APR) equation of state~\cite{Akmal:1998cf}, we obtain $R_{\rm NS} = 11.43~{\rm km}$ for $M_{\rm NS} = 1.4~M_{\odot}$. } 
\begin{align}
  \sigma_{\rm th}^{(N)} \equiv \frac{\pi R_{\rm NS}^2 m_N}{M_{\rm NS}} 
\simeq 2.5 \times 10^{-45} \times 
  \biggl(\frac{R_{\rm NS}}{11.43~{\rm km}}\biggr)^2 
  \biggl(\frac{M_{\rm NS}}{1.4~M_{\odot}}\biggr)^{-1} ~{\rm cm}^2 
  ~.
  \label{eq:sigmath}
\end{align}
This estimation agrees well with more detailed calculations; \textit{e.g.}, $\sigma^{(n)}_{\mathrm{th}} \sim 1.7 \times 10^{-45}~\mathrm{cm}^2$ is obtained in Ref.~\cite{Bell:2020jou}. 
For $\sigma^{(N)} > \sigma_{\rm th}^{(N)}$, all of the DM particles accreting onto the NS are captured, whilst for $\sigma^{(N)} < \sigma_{\rm th}^{(N)}$, the capture rate is suppressed by a factor $f \equiv \sigma^{(N)} / \sigma_{\rm th}^{(N)}$. We also note that, as shown in Ref.~\cite{Acevedo:2019agu}, the DM-nucleon scattering occurs predominantly in the NS crust for $\sigma^{(N)} \gtrsim 10^{-42}~{\rm cm}^2$. 

As we will show in Sec.~\ref{sec:scattering}, for electroweak multiplet DM, the DM-nucleon scattering can occur via inelastic scattering processes, such as $\chi^0 + n \to \chi^- + p$, where $\chi^0$ and $\chi^-$ are the neutral and charged components in the DM multiplet, respectively. For inelastic scattering to occur, the mass difference between the incoming and outgoing components must be smaller than a certain threshold value, which is of the order of the maximal energy transfer in the scattering process in the NS---from Eq.~\eqref{eq:deltae}, we expect it to be ${\cal O}(100)$~MeV. A more precise estimation of the threshold mass difference $\Delta M_{\mathrm{max}}$ is given in Ref.~\cite{Bell:2018pkk}:
\begin{equation}
  \Delta M_{\mathrm{max}} \simeq m_n \left[ e^{-\Phi (R_{\mathrm{NS}})} - 1\right] \simeq 330~\mathrm{MeV} ~,
  \label{eq:delmmax}
\end{equation}
for $m_{\mathrm{DM}} \gg m_n$. We use this as a reference value of the threshold mass difference for the DM-nucleon inelastic scattering processes in the following discussion.\footnote{Although Eq.~\eqref{eq:delmmax} is derived assuming an inelastic scattering $\chi^0_1+n\to \chi^0_2+n$ in Ref.~\cite{Bell:2018pkk}, we assume that it is also applicable to the case of $\chi^0+n\to \chi^- + p$ in the rest of this work. However, we warn that the nucleon masses are corrected by medium effects in the neutron star and depend on the model of dense nucleonic matter\,\cite{Anzuini:2021lnv,Li:2018lpy}. This leads to a proton-neutron mass split of ${\cal O}(100)$\,MeV that bears theoretical uncertainties in both the value and the sign, and further pass down to the uncertainty of $\Delta M_{\rm max}$.\label{ft:mass_splitting}}

After the first scattering, DM is gravitationally trapped by the NS and loses its energy through subsequent scatterings. As shown in Ref.~\cite{Acevedo:2019agu}, for DM particles whose annihilation cross section is as large as $2 \times 10^{-26}~\mathrm{cm}^3/\mathrm{s}$, with which their thermal relic abundance agrees with the observed DM density, the DM capture-annihilation equilibrium is achieved for old NSs even if the DM particles are distributed over the NS volume.\footnote{The distribution of DM in a NS at later times depends on the size of the scattering cross section of DM with the NS matter. If it is large enough, DM particles settle down to thermal equilibrium after a certain relaxation time. See Refs.~\cite{Bertoni:2013bsa, Garani:2018kkd,Garani:2020wge} for detailed studies on the thermalization process. The point made in Ref.~\cite{Acevedo:2019agu} is that even if the thermalization is not realized, the DM capture-annihilation equilibrium is achieved for old NSs as long as DM particles are distributed inside the NS.  } In the capture-annihilation equilibrium, the DM annihilation rate is equal to the DM capture rate in Eq.~\eqref{eq:dndt}. By noting that each DM particle eventually gives its total energy to the NS as heat, we estimate the resultant heating luminosity (observed at the distance) as 
\begin{equation}
  L_H^{\infty} =  e^{2 \Phi (R_{\mathrm{NS}})} m_{\mathrm{DM}} \left[ \chi + (\gamma_{\mathrm{esc}} -1) \right] \frac{dN}{dt} ~,
  \label{eq:lh}
\end{equation}
where $\chi$ is the fraction of the annihilation energy transferred to heat and 
the $(\gamma_{\mathrm{esc}}-1)$ factor corresponds to the contribution of the DM kinetic energy~\cite{Baryakhtar:2017dbj}. Notice that this heating luminosity does not depend explicitly on the DM mass, since, as shown in Eq.~\eqref{eq:dndt}, the DM capture rate is inversely proportional to the DM mass.

This heating effect due to the DM energy deposit modifies the temperature evolution of NSs. In the standard NS cooling scenario~\cite{Yakovlev:1999sk, Yakovlev:2000jp, Yakovlev:2004iq, Page:2004fy, Page:2009fu, Potekhin:2015qsa}, a NS cools through the emission of neutrinos and photons. It turns out that the photon emission dominates the neutrino emission for the NS age $\gtrsim 10^5$~years. The luminosity of photon emission is given by 
\begin{equation}
  L_\gamma = 4\pi R_{\mathrm{NS}}^2 \sigma_{\mathrm{SB}} T_s^4 ~,
  \label{eq:lamgam}
\end{equation}
where $\sigma_{\mathrm{SB}}$ is the Stefan-Boltzmann constant and $T_s$ is the NS surface temperature. NSs then cool down to $\lesssim 10^{3}$~K after $\sim 5 \times 10^6$~years. This scenario is, overall, consistent with the temperature observations\footnote{See Ref.~\cite{tempdata} for the latest data. } of young and middle age NSs~\cite{Potekhin:2020ttj}. In the presence of a heating source, on the other hand, the photon emission luminosity balances with the heating luminosity at late times so that  
\begin{equation}
  L_H^\infty = L_\gamma^\infty \equiv e^{2 \Phi (R_{\mathrm{NS}})} L_\gamma ~.
  \label{eq:lhlgameq}
\end{equation}
This relation determines the late-time surface temperature, which is found to be $\simeq \text{a few} \times 10^3$~K. This can be probed by future observations with infrared telescopes~\cite{Baryakhtar:2017dbj}, such as the JWST~\cite{Gardner:2006ky}.

An important caveat is that the above argument relies on the standard NS cooling theory, but its applicability to old NSs is open to debate. Indeed, recent observations indicate that there are several old NSs with temperatures much higher than the prediction in the standard NS cooling theory and the DM heating scenario~\cite{Kargaltsev:2003eb, Mignani:2008jr, Durant:2011je,  Rangelov:2016syg,  Pavlov:2017eeu, Abramkin:2021fzy}.\footnote{See Ref.~\cite{Yanagi:2019vrr} for a list of such observations. } These observations imply that there exist extra heating sources~\cite{Gonzalez:2010ta} other than the DM heating: \textit{e.g.}, the rotochemical heating~\cite{Reisenegger:1994be, 1992A&A...262..131H, 1993A&A...271..187G, Fernandez:2005cg, Villain:2005ns, Petrovich:2009yh, Pi:2009eq, Gonzalez-Jimenez:2014iia, Yanagi:2019vrr}, the vortex creep heating~\cite{1984ApJ...276..325A, 1989ApJ...346..808S, 1991ApJ...381L..47V, 1993ApJ...408..186U, VanRiper:1994vp, Larson:1998it}, and the rotationally-induced deep crustal heating \cite{Gusakov:2015kaa}. Such an extra heating source may conceal the DM heating effect. Although there is no conclusive answer to this issue for the moment, we note in passing that depending on the heating mechanism, the DM heating effect can still be observed even in the presence of extra heating sources. For example, it is shown in Ref.~\cite{Hamaguchi:2019oev} that in the presence of rotochemical heating, the DM heating effect can still be observed in ordinary pulsars with an initial period $\gtrsim 10$~ms. In this work, we simply assume that we can detect the signature of the DM heating in nearby NSs in future observations, deferring the subtle issues regarding the extra heating sources to future work. On the other hand, if a NS colder than $\sim 10^3$\,K is discovered in the future, one can rule out the extra heating effects for this star and set a limit to the DM-nucleon scattering cross section.

\section{Electroweak Multiplet Dark Matter}
\label{eq:EMDM}

The electroweak multiplet DM resides in an SU(2)$_L$ $n$-tuplet with hypercharge $Y$, where $Y$ is chosen such that there is an electrically neutral component in the multiplet; for an even number of $n$, this condition requires $Y$ to be a half-integer, while for an odd $n$, $Y$ should be an integer, with $|Y| \leq j \equiv (n-1)/2$ for both cases. As we mentioned in the introduction, we focus on the fermionic DM case for simplicity.\footnote{For discussions on the scalar and vector electroweak multiplet DM, see Refs.~\cite{Cirelli:2005uq, Cirelli:2007xd, Cirelli:2009uv, Hambye:2009pw, DiLuzio:2015oha, Logan:2016ivc, Cai:2017fmr, Belyaev:2018xpf, Chao:2018xwz, Chiang:2020rcv, Abe:2020mph,Belyaev:2020wok}. } We further divide the electroweak multiplet DM candidates into two classes---those with $Y = 0$ or $Y \neq 0$---and discuss these cases separately in Sec.~\ref{sec:yzero} and Sec.~\ref{sec:ynonzero}, respectively. For the latter case, we introduce a pair of Weyl fermions of the same SU(2)$_L$ representation with hypercharge $\pm Y$ to avoid the gauge anomalies; we label such a pair by a positive $Y$, keeping in mind that it is accompanied by a multiplet with hypercharge $-Y$. In the case of $Y =0$, we introduce only one Weyl fermion just for minimality.

A common feature of the electroweak multiplet DM is that all of the components in the multiplet have an identical mass, $M$, at tree level and the mass splittings among the components are induced by electroweak radiative corrections after the electroweak symmetry is spontaneously broken~\cite{Mizuta:1992ja, Pierce:1994ew, Cheng:1998hc, Feng:1999fu, Gherghetta:1999sw, Cirelli:2005uq}. In particular, the mass difference between the neutral component, which corresponds to the DM,
and the charged component with the electric charge $Q_m$ is induced at the one-loop level\footnote{Two-loop corrections to the mass splittings can be found in Refs.~\cite{Yamada:2009ve, Ibe:2012sx, McKay:2017xlc}. These corrections are ${\cal O}(1)$~MeV and thus negligible in the following discussions. } as 
\begin{align}
  \Delta M_m \bigr|_{\rm EW} &=
  \frac{\alpha_2}{4\pi} Q_m M
  \biggl[
  (Q_m-2Y)f\biggl(\frac{m_W}{M}\biggr)
  -(Q_m \cos^2\theta_W -2Y )f\biggl(\frac{m_Z}{M}\biggr)
  \biggr]~,
  \label{eq:delmm}
\end{align}
where $m$ denotes the label of the components in the multiplet, $\alpha_2 \equiv g_2^2/(4\pi)$ with $g_2$ the SU(2)$_L$ gauge coupling constant, $m_W$ and $m_Z$ are the masses of the $W$ and $Z$ bosons, respectively, $\theta_W$ is the weak mixing angle, and 
\begin{equation}
  f(x)\equiv -x^2+x^4\ln (x) +4x\biggl(1+\frac{x^2}{2}\biggr)\sqrt{1-\frac{x^2}{4}}
  \tan^{-1}\biggl(\frac{2}{x}\sqrt{1-\frac{x^2}{4}}\biggr)~.
\end{equation}
In the limit $x \to 0$, this function goes as $f(x) \simeq 2 \pi x +{\cal O}(x^2)$, and thus for $M \gg  m_W, m_Z$, Eq.~\eqref{eq:delmm} leads to 
\begin{equation}
  \Delta M_m  \bigr|_{\rm EW}  \simeq \alpha_2 m_W \sin^2 \frac{\theta_W}{2} \biggl[
    Q_m^2  + \frac{2Y Q_m}{\cos \theta_W }  
    \biggr]~.
    \label{eq:delmmew}
\end{equation}
This result shows that the mass splittings generated by the electroweak radiative corrections are ${\cal O}(100)$~MeV and independent of the tree-level mass $M$ at the leading order in the expansion of $m_{W,Z}/M$. For $Y = 0$, $\Delta M_m > 0$, and thus the neutral component is always the lightest. For $Y > 0$, the component with $Q_m = -1$, if exits, becomes lighter than the neutral component; to avoid this situation, we require $Y -j > -1$, \textit{i.e.}, $n < 2 Y + 3$. On the other hand, the presence of the neutral component requires $Y \leq j$, \textit{i.e.}, $n \geq 2 Y+1$, with $Y$ a half-integer for an even $n$ and an integer for an odd $n$. These conditions are simultaneously satisfied only for the cases with $n = 2Y+1$. Although other contributions to the mass 
splitting (\textit{e.g.}, the effect from additional heavy particles as we discuss below) may make the neutral component lightest even for $n \geq 2Y +3$ in some circumstances, in the following discussions, we focus on the $n = 2 Y+1$ cases for $Y > 0$.

\subsection{$Y =0$}
\label{sec:yzero}

Let us first consider a set of Weyl fermions that form an SU(2)$_L$ multiplet with the number of components $n = 3, 5, \dots$ and zero hypercharge. We denote them by $\chi_m$ where the subscript $m = -j, -j+1, \dots, j$ with $j = (n-1)/2$ is the eigenvalue of the SU(2)$_L$ diagonal generator $T_3$. The renormalizable Lagrangian terms for these fields are then given by\footnote{We note that $(-1)^m \chi_{-m}$ transforms in the same way as the complex conjugate representation of $\chi_m$. }
\begin{equation}
  {\cal L}_{\rm ren} = \sum_{m=-j}^{j}
  \chi_m ^\dagger (i\overline{\sigma}^\mu D_\mu  \chi)_{m}
  -\frac{1}{2}\sum_{m=-j}^{j} (-1)^{m} (M \chi_{-m} \chi_{m}+
  \text{h.c.})~,
\end{equation} 
where $D_\mu$ is the covariant derivative. We take $M$ to be real and positive without loss of generality. The $m=0$ component $\chi_0$ is electrically neutral, and thus is the DM in the present case. 

Since this multiplet has only the gauge interactions, the relic abundance of $\chi_0$ is determined unambiguously as a function of the DM mass $M$. It is found that the observed DM density is reproduced with $M \simeq 3$~TeV for $n=3$~\cite{Hisano:2006nn} and with $M \simeq 14$~TeV for $n=5$~\cite{Mitridate:2017izz}.

If the theory has extra heavy states that couple to both the DM and SM fields, they induce interactions among these fields. These interactions are described by non-renormalizable effective operators at low energies with their coefficients suppressed by a power of a cut-off scale $\Lambda$, which is of the same order as the mass scale of the heavy states. Among such operators, 
\begin{equation}
  {\cal L}_5 = - \frac{c_5}{2\Lambda}  \sum_{m=-j}^{j} (-1)^{m} \chi_{-m} \chi_{m} |H|^2 + {\rm h.c.} 
  \label{eq:l5y0}
\end{equation}
have the lowest mass dimensions, where $H$ is the SM Higgs field. Notice that the operators of the form 
\begin{equation}
  {\cal L}_5^\prime = - \frac{c_5^\prime}{2\Lambda} 
  \sum_{m,n} (-1)^m \chi_{-m} (T_a)_{mn} \chi_n \, 
  H^\dagger \tau_a H +{\rm h.c.}  
  \label{eq:l5primey0}
\end{equation}
vanish, where $T_a$ and $\tau_a$ are the $n$-dimensional and fundamental representations of the SU(2)$_L$ generators, respectively.\footnote{We adopt the following convention for these matrices:
\begin{align}
  (T_1)_{mn} &= \frac{1}{2} \left[ \sqrt{(j-n)(j+n+1)} \,\delta_{m,n+1} 
  + \sqrt{(j+n)(j-n+1)} \,\delta_{m,n-1} 
  \right] ~, \nonumber \\
  (T_2)_{mn} &= \frac{1}{2i} \left[ \sqrt{(j-n)(j+n+1)}\, \delta_{m,n+1} 
  - \sqrt{(j+n)(j-n+1)} \,\delta_{m,n-1} 
  \right] ~, \nonumber \\
   (T_3)_{mn} &=  n \,\delta_{m,n} ~,
\end{align}
and $\tau_a = \sigma_a/2$ with $\sigma_a$ the Pauli matrices. 
} One can readily show this by noting that the quantity $(-1)^m (T_a)_{-m, n}$ is antisymmetric with respect to the exchange of $m$ and $n$. After the Higgs field acquires a vacuum expectation value (VEV), 
\begin{equation}
  \langle H \rangle = \frac{1}{\sqrt{2}} 
  \begin{pmatrix}
    0 \\ v 
  \end{pmatrix}
  ~,
\end{equation}
with $v \simeq 246$~GeV, the operators in Eq.~\eqref{eq:l5y0} lead to a common mass term; namely, these operators do not generate mass splittings among the components. To find the operators with the lowest mass dimensions that give rise to the neutral-charged mass splitting, we first note that such an operator has the form $C_{mn} \chi_m \chi_n$ and $C_{mn}$ should transform non-trivially under the SU(2)$_L$ transformations~\cite{Gherghetta:1999sw}. Moreover, $C_{mn}$ should be a symmetric matrix, which means that $C_{mn}$ transforms as a $j = 2, 4, 6, \dots$ representation of SU(2)$_L$.\footnote{This explains why the operator in Eq.~\eqref{eq:l5primey0} vanishes; $H^\dagger \tau_a H$ transforms as a $j=1$ representation, but this possibility has been excluded from this argument. } We can construct such a quantity by means of the Higgs field $H$, and the one with the lowest mass dimension corresponds to $j = 2$. We thus find that the following dimension-seven operators are the lowest-dimensional ones that generate mass splittings: 
\begin{equation}
  {\cal L}_7 =  - \frac{c_7}{2\Lambda^3} 
  \sum_{m,n} (-1)^m \chi_{-m} (T_a T_b)_{mn} \chi_n \, 
  (H^\dagger \tau_a H) (H^\dagger \tau_b H) +{\rm h.c.}  
  \label{eq:dim7}
\end{equation}  

After the ${\rm SU}(2)_L \times {\rm U}(1)_Y$ symmetry is spontaneously broken, the tree-level masses of the neutral ($m=0$) and charged ($m = \pm 1$) components are then given respectively by
\begin{align}
  M_0 &= M + \frac{c_5v^2}{2\Lambda} ~, \\[2pt]
  M_{\pm} &= M + \frac{c_5v^2}{2\Lambda}  + \frac{c_7 v^4}{16\Lambda^3} ~.
\end{align}
By taking account of the electroweak radiative corrections in Eq.~\eqref{eq:delmm} as well, we obtain the neutral-charged mass splitting as 
\begin{equation}
  \Delta M_{\pm} = \frac{c_7 v^4}{16 \Lambda^3} + \Delta M_{\pm} \bigr|_{\rm EW} ~,
  \label{eq:delmplmiy0}
\end{equation}
where at the one-loop level
\begin{align}
  \Delta M_{\pm} \bigr|_{\rm EW} &=
  \frac{\alpha_2}{4\pi}  M
  \biggl[
  f\biggl(\frac{m_W}{M}\biggr)
  -\cos^2\theta_W f\biggl(\frac{m_Z}{M}\biggr)
  \biggr]~.
  \label{eq:delmpmy0}
\end{align}
We find that this expression is the same for $n = 3, 5, \dots$. The two-loop calculation in Ref.~\cite{McKay:2017xlc} shows $\Delta M_{\pm} |_{\rm EW} \simeq 165$~MeV for $M \gtrsim 1$~TeV for $n = 3, 5$. On the other hand, the first term in the right-hand side of Eq.~\eqref{eq:delmplmiy0} is computed as 
\begin{equation}
  \frac{c_7 v^4}{16 \Lambda^3} = c_7 \, \biggl(\frac{\Lambda}{10~{\rm TeV}}\biggr)^{-3} \times 0.23~{\rm MeV} ~.
\end{equation}
Consequently, the contribution of the dimension-seven operator~\eqref{eq:dim7} is much smaller than the electroweak radiative corrections for $\Lambda \gg 1$~TeV. 

Generically speaking, the coefficient $c_5$ in Eq.~\eqref{eq:l5y0} is a complex parameter. A non-zero complex phase in $c_5$ induces the electric dipole moments (EDMs) of quarks and leptons via the two-loop Barr-Zee diagrams~\cite{Barr:1990vd}, which are stringently constrained by experiments. In particular, the current limit on the electron EDM imposes $\Lambda/|{\rm Im} (c_5)| \gtrsim {\cal O}(10)$~TeV~\cite{Hisano:2014kua, Nagata:2014wma, Nagata:2014aoa,  Kuramoto:2019yvj}. Considering this, in what follows, we always take the coefficients of the effective operators to be real to evade the EDM limits and thus to obtain a conservative limit on the cut-off scale $\Lambda$.

\subsection{$Y \neq 0$}
\label{sec:ynonzero}

Next, we consider the DM multiplets with non-zero hypercharge~\cite{Nagata:2014aoa}. As we discussed above, we focus on the cases with $n = 2Y +1$ ($Y > 0$), \textit{i.e.}, $j=Y$. We introduce a pair of Weyl fermions, $\chi_m$ and $\eta_m$, which transform as the same SU(2)$_L$ representation and have hypercharge $Y$ and $-Y$, respectively. The renormalizable Lagrangian terms in this case are then given by 
\begin{equation}
  {\cal L}_{\rm ren} = \sum_{m=-j}^{j}
  \chi_m ^\dagger (i\overline{\sigma}^\mu D_\mu  \chi)_{m}
  +\sum_{m=-j}^{j}
  \eta_m ^\dagger (i\overline{\sigma}^\mu D_\mu  \eta)_{m}
  -\sum_{m=-j}^{j} (-1)^{j+m} (M\eta_{-m} \chi_{m}+
  \text{h.c.})~.
  \label{eq:lrenhyp}
\end{equation} 
The neutral components in this case are $\chi_{-Y}$ and $\eta_{+Y}$, which have an identical mass and thus form a Dirac fermion if only the terms in Eq.~\eqref{eq:lrenhyp} are taken into account.
The thermal relic abundance of these neutral components is found to agree with the observed DM density if their mass is $\simeq 1.1$~TeV, $\simeq 1.9$~TeV, and $\simeq 2.6$~TeV for $n = 2$ and $Y = 1/2$, $n = 3$ and $Y = 1$, and $n = 4$ and $Y =3/2$, respectively~\cite{Farina:2013mla}.  

In the present case, the dimension-five effective operators that couple the DM multiplet with the Higgs field are 
\begin{equation}
  {\cal L}_5 = - \frac{c_5}{\Lambda}  \sum_{m} (-1)^{j+m} \eta_{-m} \chi_{m} |H|^2 
  - \frac{c_5^\prime}{\Lambda} 
  \sum_{m,n} (-1)^{j+m} \eta_{-m} (T_a)_{mn} \chi_n \, 
  H^\dagger \tau_a H 
  + {\rm h.c.} 
  \label{eq:l5ynon0}
\end{equation}
After the Higgs field develops a VEV, the second operator in the above equation generates a mass splitting between the neutral and charged components.

In addition, there exist dipole operators for the DM multiplet, which are also of dimension-five. We, however, do not consider these operators in the following discussions; firstly, they do not generate mass splittings among the components in the DM multiplet, and thus are irrelevant in the discussion on the inelastic DM scattering in NSs. Secondly, although these operators do contribute to the DM-nucleon scattering \cite{Bagnasco:1993st, Sigurdson:2004zp, Masso:2009mu, Cho:2010br, Banks:2010eh}, these operators are often induced at loop level \cite{Ibarra:2015fqa, Herrero-Garcia:2018koq, Hisano:2018bpz} and thus their contribution to the scattering cross section is subdominant compared with the other contributions. 

The terms in Eqs.~\eqref{eq:lrenhyp} and \eqref{eq:l5ynon0} have a global U(1) symmetry associated with the DM number: 
\begin{equation}
  \chi_m \to e^{i\theta}\chi_m~, \qquad
  \eta_{-m} \to e^{-i \theta} \eta_{-m} ~. 
  \label{eq:u1sym}
\end{equation}
As a result, each component still forms a Dirac fermion, and in particular the neutral components $\chi_{-Y}$ and $\eta_{+Y}$ are degenerate in mass. This is, however, problematic since the SI scattering cross section of hypercharged Dirac DM with nucleons is so large due to the presence of the vector coupling with $Z$ boson that it has already been excluded by DM direct detection experiments. This problem is evaded if there are some interactions that break the symmetry~\eqref{eq:u1sym},\footnote{Another way of avoiding this problem is to take the DM mass to be as large as $10^8$~GeV~\cite{Feldstein:2013uha}. In this case, the DM should not be thermalized in the early Universe and would be produced non-thermally at the reheating epoch. We do not further consider this possibility in this paper.  } with which the Dirac fermion DM is decomposed into two Majorana fermions with different masses. The lowest-dimensional operators that describe such interactions are~\cite{Nagata:2014aoa} 
\begin{align}
  \mathcal{L}_{\text{spl}}= &-\frac{c_s}{2\Lambda^{(4Y-1)}}
 \sum_{M,m,m^\prime}\langle jmjm^\prime |(2Y)M\rangle  [(H)^{4Y}_M ]^*
 \chi_m \chi_{m^\prime}
 \nonumber \\&-\frac{c_s^\prime}{2\Lambda^{(4Y-1)}}
 \sum_{M,m,m^\prime} (-1)^{2Y +M}
 \langle jmjm^\prime |(2Y)M\rangle  [(H)^{4Y}_{-M} ]\,
 \eta_m \eta_{m^\prime}
+\text{h.c.}~,
 \label{eq:effc}
\end{align}
where $(H)^k$ is composed of $k$ Higgs fields to form an isospin-$k/2$
object and defined such that its lowest component is given
by $(H^0)^k$, and $\langle j mj^\prime m^\prime |JM\rangle $ are the
Clebsch-Gordan coefficients. Notice that $\langle jmjm^\prime |(2Y)M\rangle = \langle jm^\prime jm|(2Y)M\rangle$ for $j-Y = 0$. 

A non-zero VEV of the Higgs field generates the following mass matrix for the neutral components: 
\begin{equation}
  {\cal L}_{\rm mass} = - \frac{1}{2} (\chi_{-Y}, \eta_{+Y}) \, {\cal M}
  \begin{pmatrix}
    \chi_{-Y} \\ \eta_{+Y}
  \end{pmatrix}
  +{\rm h.c.}
  ~,
\end{equation}
with 
\begin{equation}
  {\cal M} = 
  \begin{pmatrix}
    \frac{c_s  v^{4Y}}{2^{2Y}\Lambda^{(4Y-1)}}  & M + \frac{c_5 v^2}{2\Lambda} + \frac{c_5^\prime Y v^2}{4\Lambda}\\[3pt]
    M + \frac{c_5 v^2}{2\Lambda} + \frac{c_5^\prime Y v^2}{4\Lambda}& \frac{c_s^\prime  v^{4Y}}{2^{2Y}\Lambda^{(4Y-1)}} 
  \end{pmatrix}
  ~,
\end{equation}
where we have used $\langle j j j j |(2j) (2j)\rangle = 1$.\footnote{Note that $\langle j (-j) j (-j) |(2j) (-2j)\rangle  = \langle j j j j |(2j)(2j)\rangle$.} This mass matrix is diagonalized with a unitary matrix $U$ as 
\begin{equation}
  U^* {\cal M} U^\dagger =
  \begin{pmatrix}
    M_{\chi_1^0} & 0 \\ 0 & M_{\chi_2^0} 
  \end{pmatrix}
  ~,
\end{equation}
where
\begin{align}
  M_{\chi_{1,2}^0} \simeq M + \frac{c_5 v^2}{2\Lambda} + \frac{c_5^\prime Y v^2}{4\Lambda} 
  \mp \frac{v^{4Y} |c_s + c_s^\prime|}{2^{2Y+1} \Lambda^{(4Y-1)}} ~.
\end{align}
The mass difference between the neutral components is thus found to be 
\begin{equation}
  \Delta M_0 \simeq \frac{v^{4Y} |c_s + c_s^\prime|}{2^{2Y} \Lambda^{(4Y-1)}} ~. 
  \label{eq:delm0}
\end{equation}
The eigenstate with the smaller mass eigenvalue, $M_{\chi_{1}^0}$, corresponds to the DM in this model. It is described by a Majorana fermion and therefore does not have the vector coupling with $Z$ boson. As a result, the SI elastic scattering via the $Z$-boson exchange vanishes in the presence of the operators~\eqref{eq:effc}. On the other hand, the SI inelastic scattering process accompanied by the heavier neutral state through the $Z$-boson exchange can still occur, if the mass splitting $\Delta M_0$ is smaller than the energy transfer in the scattering process. In order to evade severe constraints from DM direct detection experiments, $\Delta M_0 $ is required to be larger than a few hundreds of keV. 

The mass difference between the charged component and DM is 
\begin{equation}
  \Delta M_\pm \simeq - \frac{c_5^\prime v^2}{4\Lambda} +
  \frac{v^{4Y} |c_s+c_s^\prime |}{2^{2Y+1} \Lambda^{(4Y-1)}} 
  +\Delta M_{\pm} \bigr|_{\rm EW}
  ~,
  \label{eq:delmplmiynon0}
\end{equation}
where at the one-loop level
\begin{align}
  \Delta M_{\pm} \bigr|_{\rm EW} &=
  \frac{\alpha_2}{4\pi} M
  \biggl[
  (1-2Y)f\biggl(\frac{m_W}{M}\biggr)
  -(\cos^2\theta_W -2Y )f\biggl(\frac{m_Z}{M}\biggr)
  \biggr]~.
  \label{eq:delmpmynon0eq}
\end{align}
We note that the first term in the right-hand side of Eq.~\eqref{eq:delmplmiynon0} is suppressed only by a factor of $\Lambda$, contrary to the case of $Y=0$ where the contribution of higher-dimensional operators to the neutral-charged mass splitting is suppressed at least by a factor of $\Lambda^3$, as seen in Eq.~\eqref{eq:delmplmiy0}.

As we discussed above, some effect that induces the mass splitting between the neutral components, such as those in Eq.~\eqref{eq:effc}, is required to evade the limits from direct detection experiments. Generically, this effect induces also the lower-dimensional operators like those in Eq.~\eqref{eq:l5ynon0} unless they are forbidden by symmetry; namely, if we normalize the cut-off scale $\Lambda$ in Eq.~\eqref{eq:effc} such that $c_s, c_s^\prime$ are ${\cal O}(1)$, then we expect $c_5$ and $c_5^\prime$ are also ${\cal O}(1)$. We will assume this in the following analysis.

\begin{figure}[t]
  \centering
  {\includegraphics[width=0.55\textwidth]{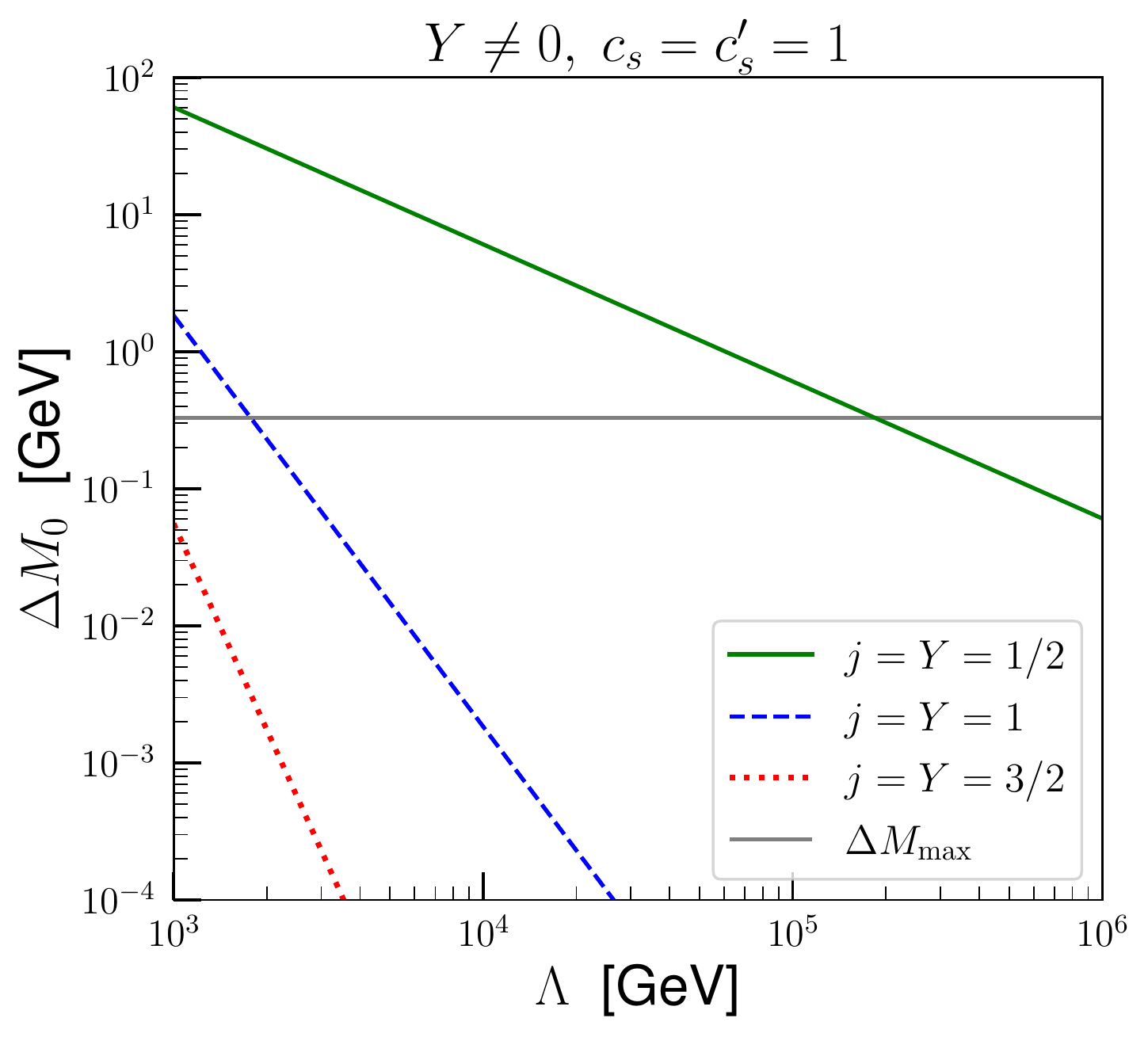}} 
  \caption{
   The mass difference between the neutral components, $\Delta M_0$, as a function of the cut-off scale $\Lambda$ {for $Y\ne 0$,} where we take $c_s = c_s^\prime = 1$. The green solid, blue dashed, and red dotted lines correspond to the cases with $j = Y = 1/2$, $1$, and $3/2$, respectively. The horizontal line indicates the threshold mass difference $\Delta M_{\mathrm{max}}$ in Eq.~\eqref{eq:delmmax}. 
  }
  \label{fig:delm0}
\end{figure}

In Fig.~\ref{fig:delm0}, we show the mass difference between the neutral components, $\Delta M_0$ in Eq.~\eqref{eq:delm0}, as a function of the cut-off scale $\Lambda$, where we take $c_s = c_s^\prime = 1$. The green solid, blue dashed, and red dotted lines correspond to the cases with $j = Y = 1/2$, $1$, and $3/2$, respectively. This figure shows that for the $j\geq 1$ cases, the requirement $\Delta M_0 \gtrsim \text{a few} \times 100$~keV, which is required to evade the direct detection limits, gives a strong limit on the cut-off scale $\Lambda$~\cite{Nagata:2014aoa}: $\Lambda \lesssim \text{a few} \times 10^{4}$~GeV and a few TeV for $j= 1$ and $3/2$, respectively. For the doublet case, this bound is quite weak. We also find that in a wide range of $\Lambda$, the mass splitting is predicted to be smaller than the threshold mass difference $\Delta M_{\mathrm{max}}$ in Eq.~\eqref{eq:delmmax}, which is indicated by the horizontal line in Fig.~\ref{fig:delm0}. Even for $j = 1/2$, $\Delta M_0 \lesssim \Delta M_{\mathrm{max}}$ is obtained for $\Lambda \gtrsim 2 \times 10^5$~GeV.

\begin{figure}[t]
  \centering
  {\includegraphics[width=0.9\textwidth]{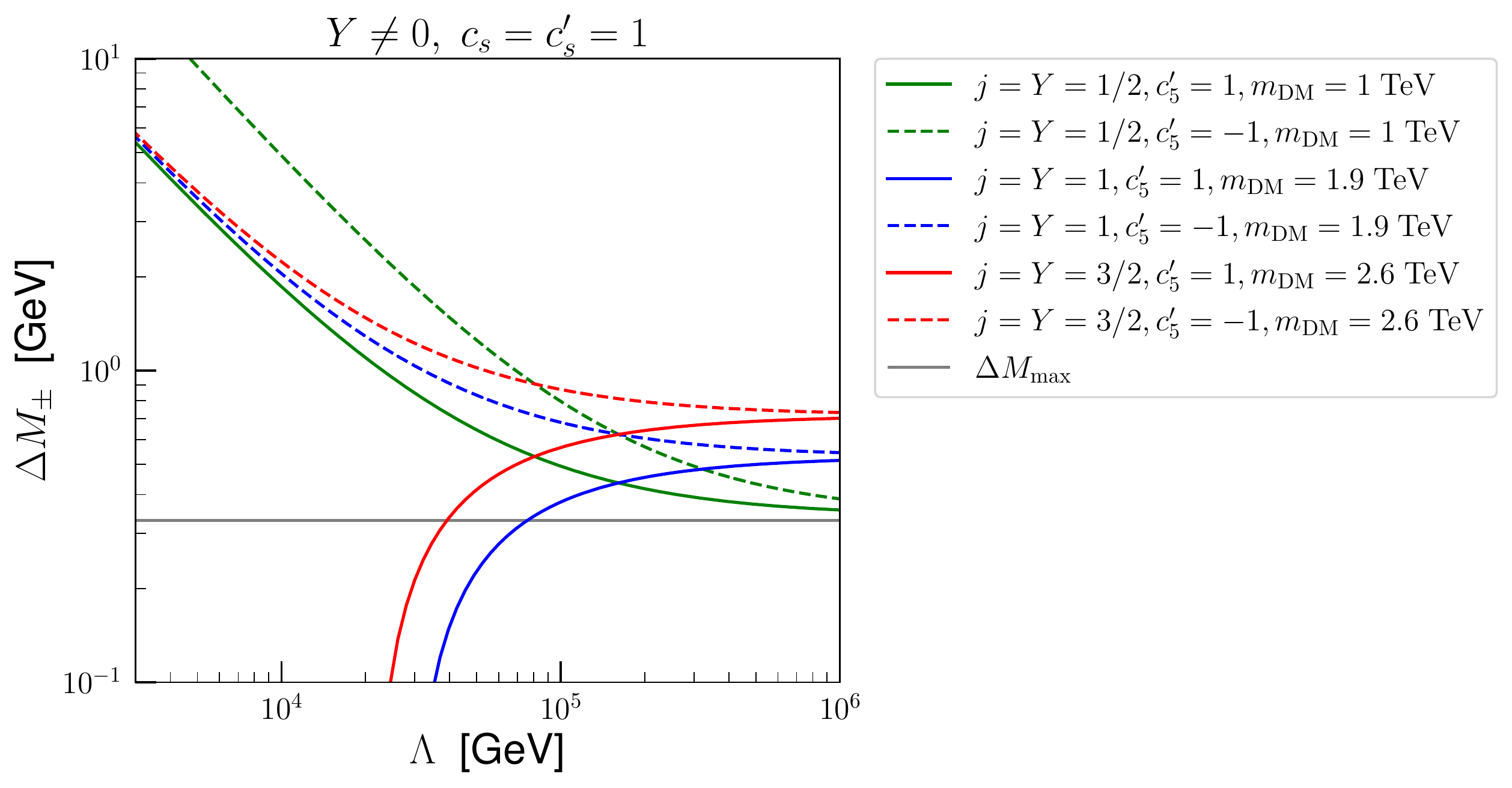}} 
  \caption{
   The mass difference between the charged component and the DM, $\Delta M_{\pm}$, as a function of the cut-off scale $\Lambda$
   {for $Y\ne 0$}, where we take $c_s = c_s^\prime = 1$. The green, blue, and red lines correspond to the cases with $j = Y = 1/2$, $1$, and $3/2$, respectively, and the solid (dashed) lines represent the cases with $c_5^\prime = + 1$ ($-1$). We fix the DM mass, $m_{\rm DM}$, to be 1~TeV, 1.9~TeV, and 2.6~TeV for the doublet, triplet, and quartet cases, respectively, with which the observed DM density can be explained by the thermal relic abundance. The horizontal line indicates the threshold mass difference $\Delta M_{\mathrm{max}}$ in Eq.~\eqref{eq:delmmax}. 
  }
  \label{fig:delmpm}
\end{figure}

Next, we show in Fig.~\ref{fig:delmpm} the mass difference between the charged component and the DM, $\Delta M_{\pm}$ in Eq.~\eqref{eq:delmplmiynon0}, as a function of the cut-off scale $\Lambda$, where we take $c_s = c_s^\prime = 1$. The green, blue, and red lines correspond to the cases with $j = Y = 1/2$, $1$, and $3/2$, respectively, and the solid (dashed) lines represent the cases with $c_5^\prime = + 1$ ($-1$). We fix the DM mass, $m_{\rm DM}$, to be 1~TeV, 1.9~TeV, and 2.6~TeV for the doublet, triplet, and quartet cases, respectively, with which the observed DM density can be explained by the thermal relic abundance in the large $\Lambda$ limit. The horizontal line again indicates the threshold mass difference $\Delta M_{\mathrm{max}}$ in Eq.~\eqref{eq:delmmax}.  As we see, all lines converge to asymptotic values in the large $\Lambda$ limit, which correspond to $\Delta M_{\pm}|_{\rm EW}$; the asymptotic value is larger for a larger $Y$, as can be seen from Eq.~\eqref{eq:delmmew}. The behavior of $\Delta M_{\pm}$ in the lower $\Lambda$ region depends on $Y$ and ${\rm sgn}(c_5^\prime)$. For $c_5^\prime = +1$ and $Y = 1$ or $3/2$, $\Delta M_{\pm}$ rapidly decreases as $\Lambda$ gets smaller, and eventually becomes negative when the first term dominates the third term in Eq.~\eqref{eq:delmplmiynon0}---this case is phenomenology disfavored as the charged state becomes the lightest. For the other cases, $\Delta M_{\pm}$ increases in the lower $\Lambda$ region.\footnote{We, however, note that this behavior depends on the relative size between $c_5^\prime$ and $|c_s + c_s^\prime|$ for the doublet case; for $c_5^\prime > |c_s + c_s^\prime|$, $\Delta M_{\pm}$ would decrease as $\Lambda$ gets smaller. } In any case, it turns out that the mass splitting $\Delta M_{\pm}$ tends to be larger than the threshold mass difference $\Delta M_{\mathrm{max}}$, unless there is cancellation among the terms in Eq.~\eqref{eq:delmplmiynon0}.

\section{Dark Matter-Nucleon Scattering}
\label{sec:scattering}

Next, we discuss the DM-nucleon scattering both in a NS and on the earth---\textit{i.e.}, in DM direct detection experiments. The DM-nucleon scattering processes are classified into two categories: the SI and spin-dependent (SD) scattering processes. The significance of each scattering process differs in NSs and the earth. Firstly, in DM direct detection experiments, the SI scattering processes are more important than the SD ones since the cross section of the former is enhanced by coherent scattering over the target nuclei, while there is no such enhancement in the latter. Secondly, inelastic scattering processes are prohibited in direct detection experiments if the mass splittings $\Delta M_{\pm}$ and $\Delta M_0$ are larger than a few hundreds of keV. In a NS, on the other hand, when DM scatters with the NS matter, its velocity is as large as $\gtrsim 0.5 c$ 
and, as a result, the energy transfer in the scattering tends to be much larger than that on the earth. This enables inelastic scattering processes to occur even for $\Delta M_{\pm}, \Delta M_0 \simeq {\cal O}(100)$~MeV, as we have seen in Sec.~\ref{sec:heating}. Thirdly, DM scatters mainly with nucleons in NSs, not with nuclei as in direct detection experiments, and thus there is no difference in the significance of the SD and SI scatterings. 

\begin{table}[t]
  \renewcommand{\arraystretch}{2}
  \centering
  \caption{
  {Reference values of the DM-nucleon scattering cross sections and mass splitting thresholds that determine the detectability of the TeV-scale DM in DM direct detection experiments and NS observation. 
  }
  }
  \begin{tabular}{ c  | c  | c  | c  }
  \hline  
  \hline
  \multicolumn{2}{c|}{}  &  Direct detection  &  NS observation
  \\  
  \hline
  \multirow{2}{*}{Elastic}  
  &  SI  
  &  {$\sigma_{\rm  SI}^{(N), {\rm  upper}}  \simeq  10^{-45}~\mathrm{cm}^2$}
  &  {\multirow{2}{*}{$\sigma_{\rm  th}^{(N)}  \simeq 10^{-45}~\mathrm{cm}^2$}}
  \\  
  \cline{2-3}
  &  SD  
  & {$\sigma_{\rm  SD}^{(N), {\rm  upper}}    \simeq  10^{-40}~\mathrm{cm}^2$}
  &  
  \\  
  \hline
  \multirow{2}{*}{Inelastic}  
  &  SI  
  &  \multirow{2}{*}{{$\Delta  M_0,  \Delta  M_{\pm}  \lesssim  {\cal  O}  (100)~\mathrm{keV}$}}  
  &  \multirow{2}{*}{{$\Delta  M_0,  \Delta  M_{\pm} \lesssim  {\cal  O} (100)~\mathrm{MeV}$}}  
  \\  
  \cline{2-2}
  &  SD  &  &
  \\  
  \hline
  \hline
  \end{tabular}
  \label{tab:DDvsNS}
\end{table}

To compare the sensitivities of DM direct detection experiments and NS observation to the DM-nucleon scattering processes, in Table~\ref{tab:DDvsNS}, we show the reference values of the cross sections and mass splitting thresholds that determine the detectability of DM for each case. For elastic scattering in direct detection experiments, we show the current upper limits on its cross section for the TeV-scale DM; for the SI scattering, the PandaX-4T experiment gives $\sigma_{\rm  SI}^{(N), {\rm  upper}}  \simeq  10^{-45}~\mathrm{cm}^2$~\cite{PandaX-4T:2021bab} while for the SD case, the XENON1T~\cite{XENON:2019rxp} and PICO-60~\cite{PICO:2019vsc} experiments give $\sigma_{\rm  SD}^{(N), {\rm  upper}}  \simeq  10^{-40}~\mathrm{cm}^2$ for neutron and proton, respectively. For elastic scattering in NSs, on the other hand, the relevant quantity is the threshold  cross section~\eqref{eq:sigmath}, $\sigma_{\rm  th}^{(N)}  \simeq  10^{-45}~\mathrm{cm}^2$. This is comparable to the current limit for the SI scattering but much lower than that for the SD scattering. For inelastic scattering, the threshold mass difference determines whether this process occurs or not; this is $\simeq \text{a few} \times 100 ~\mathrm{keV}$ in direct detection experiments and $ \Delta M_{\mathrm{max}}\simeq \text{a few} \times 100 ~\mathrm{MeV}$ for NSs (see Eq.~\eqref{eq:delmmax}).

In what follows, we compute the cross sections of these DM nucleon scattering processes, {\it i.e.}, the SD/SI elastic/inelastic scatterings. We, again, discuss the $Y=0$ and $Y \neq 0$ cases separately in Sec.~\ref{sec:yzerodn} and Sec.~\ref{sec:ynonzerodn}, respectively.

\subsection{$Y = 0$}
\label{sec:yzerodn}

\subsubsection{Elastic scattering}

We first consider the DM-nucleon elastic scattering process for the $Y=0$ case. Let us begin with the SI scattering. The SI scattering cross section of Majorana fermion DM with a nucleon $N$ is given by 
\begin{equation}
  \sigma_{\rm SI}^{(N)} = \frac{4}{\pi} \biggl(\frac{m_{\rm DM} m_N}{m_{\rm DM} + m_N}\biggr)^2 \, f_N^2 ~,
  \label{eq:sigmasi}
\end{equation}
where $m_N$ is the nucleon mass and $f_N$ is the effective DM-nucleon SI coupling. There are two kinds of processes that contribute to this effective coupling. First, the dimension-five operator \eqref{eq:l5y0} induces the DM-nucleon scattering via the Higgs-boson exchange, with the effective coupling given by~\cite{Shifman:1978zn}
\begin{equation}
  f_{N}^{(5)} = \frac{c_5 m_N}{2\Lambda m_h^2} \biggl[\sum_{q = u,d,s} f_{T_q}^{(N)} + 3 \times \frac{2}{27} f_{T_G}^{(N)} \biggr] ~,
\end{equation}
where $m_h$ is the Higgs-boson mass, $f_{T_q}^{(N)} \equiv \langle N | m_q \bar{q} q| N \rangle/m_N$ are the nucleon matrix elements of the quark scalar operators, and $f_{T_G}^{(N)} \equiv 1- \sum_{q = u,d,s} f_{T_q}^{(N)}$. For the nucleon matrix elements $f_{T_q}^{(N)}$, we use the values obtained by a recent compilation~\cite{Ellis:2018dmb}: $f^{(p)}_{T_u} = 0.018$, $f^{(p)}_{T_d} = 0.027$, $f^{(p)}_{T_s} = 0.037$, $f_{T_u}^{(n)} = 0.013$, $f_{T_d}^{(n)} = 0.040$, $f_{T_s}^{(n)} = 0.037$. 

Second, the DM-nucleon scattering is also induced via $W$-boson loop processes~\cite{Hisano:2004pv, Hisano:2010fy, Hisano:2010ct, Hisano:2011cs, Hill:2013hoa, Hill:2014yxa, Hisano:2015rsa}. This contribution to the effective coupling $f_N$ is expressed as 
\begin{equation}
  f_N^{(\rm EW)} =  (n^2 -1) f_N^W ~. 
  \label{eq:fnewy0}
\end{equation}
We use the results given in Ref.~\cite{Hisano:2015rsa} to compute $f_N^W$. It is found that if $m_{\rm DM} \gg m_W$, $f_N^W$ rarely depends on the DM mass, having a constant value:\footnote{This non-decoupling behavior results from the properties of the loop integrals for these processes; they are dominated by the contribution from the energy scale of $\mathcal{O}(m_W)$ and thus remain non-vanishing even if the DM mass is very large~\cite{Hisano:2004pv}. } 
\begin{equation}
  f_p^W \simeq 2.8 \times 10^{-11} ~{\rm GeV}^{-2} ~, \qquad 
  f_n^W \simeq 2.7 \times 10^{-11} ~{\rm GeV}^{-2} ~.
\end{equation}

\begin{figure}
  \centering
  \subcaptionbox{\label{fig:sigsi3}
  $n = 3$
  }
  {\includegraphics[width=0.48\textwidth]{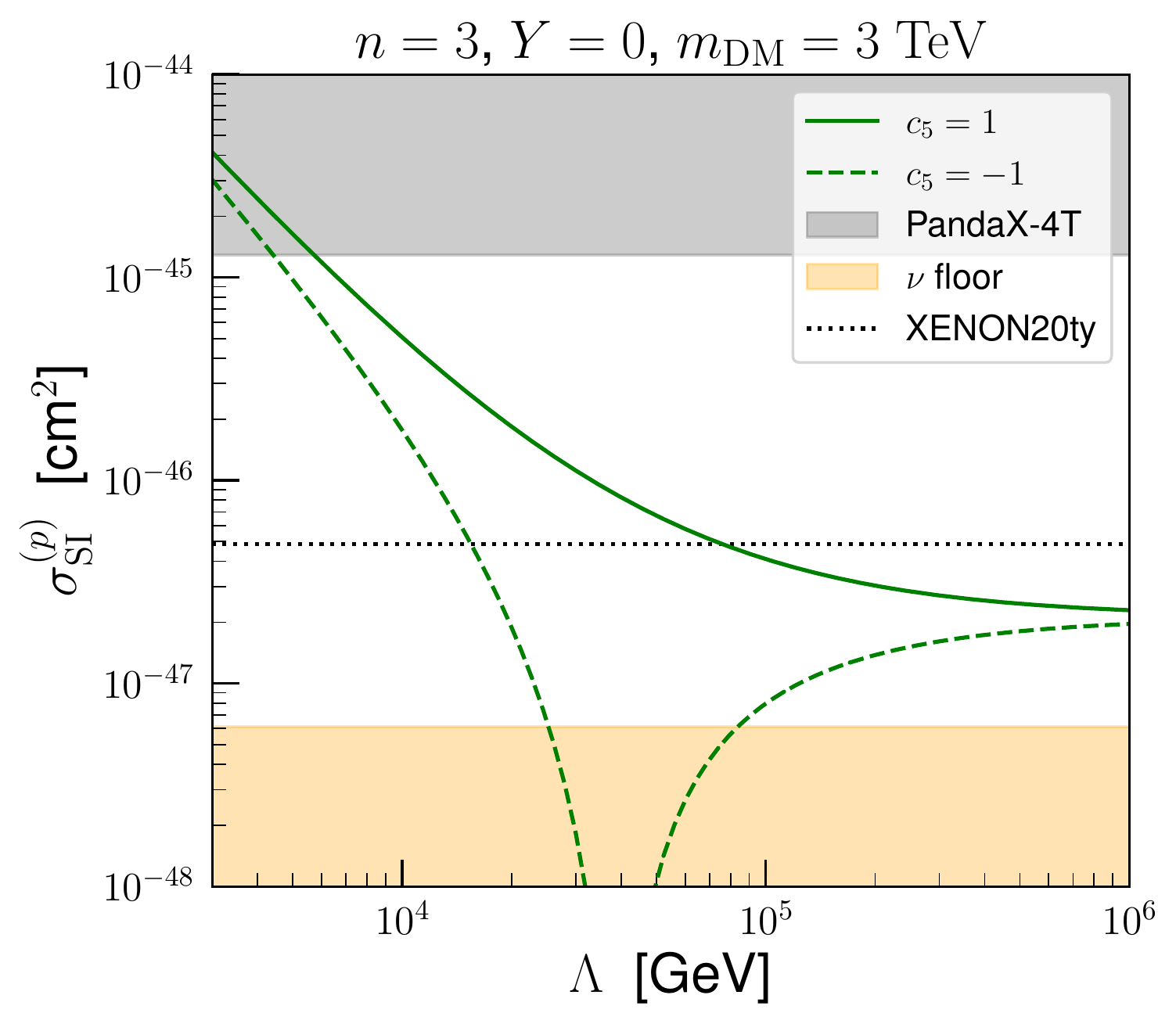}}
  \subcaptionbox{\label{fig:sigsi5}
  $n = 5$
  }
  { 
  \includegraphics[width=0.48\textwidth]{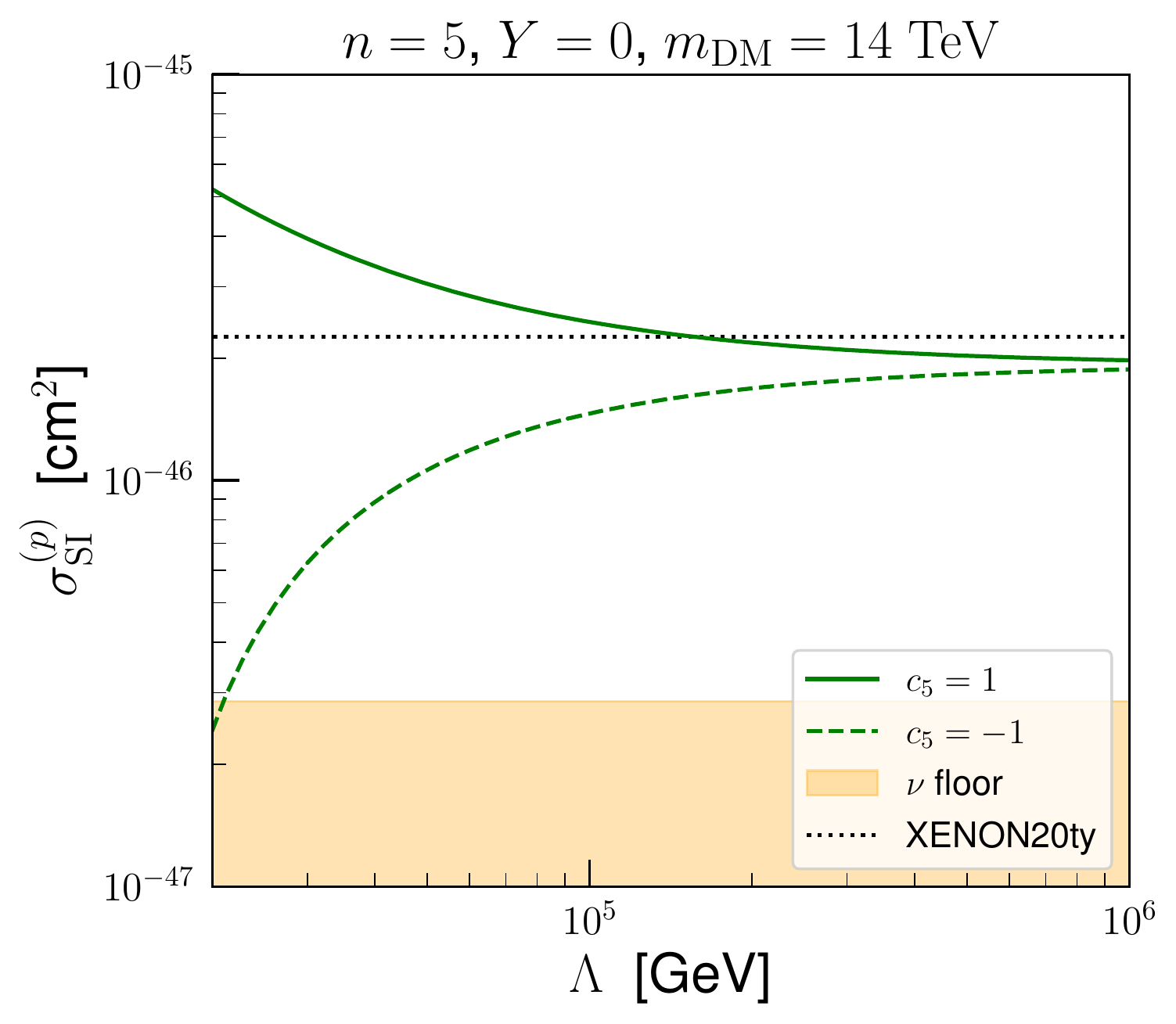}}
  \caption{
    The SI DM-proton elastic scattering cross section {for $Y=0$,} as a function of the cut-off scale $\Lambda$ for (a) $n = 3$ and $m_{\rm DM} = 3$~TeV and (b) $n=5$ and $m_{\rm DM} = 14$~TeV. The green solid and dashed lines correspond to the cases with $c_5 = +1$ and $-1$, respectively. The gray shaded area is excluded by the PandaX-4T experiment~\cite{PandaX-4T:2021bab}. The horizontal dotted line represents the expected sensitivity of the XENONnT experiment with a 20 ton-year exposure~\cite{XENON:2020kmp}, and the orange area corresponds to the neutrino floor~\cite{Billard:2021uyg}.
  }
  \label{fig:sicrossy0}
  \end{figure}

The effective coupling $f_N$ is then given by the sum of these two contributions: $f_N = f_N^{(5)} + f_N^{(\rm EW)}$. Using this effective coupling and Eq.~\eqref{eq:sigmasi}, we compute the SI DM-proton elastic scattering cross section, which is shown in Fig.~\ref{fig:sicrossy0} as a function of the cut-off scale $\Lambda$ for $n = 3$ and $m_{\rm DM} = 3$~TeV (Fig.~\ref{fig:sigsi3}) and $n=5$ and $m_{\rm DM} = 14$~TeV (Fig.~\ref{fig:sigsi5}). The green solid and dashed lines correspond to the cases with $c_5 = +1$ and $-1$, respectively. The gray shaded area is excluded by the PandaX-4T experiment~\cite{PandaX-4T:2021bab}. The horizontal dotted line represents the expected sensitivity of the XENONnT experiment with a 20 ton-year exposure~\cite{XENON:2020kmp}, and the orange area corresponds to the neutrino floor~\cite{Billard:2021uyg}. For both $n = 3, 5$, the SI cross sections approach certain values in the limit of large $\Lambda$, which correspond to the electroweak loop contribution. These values are out of the reach of the XENONnT experiment, but well above the neutrino floor. The behavior of the SI cross sections for lower values of $\Lambda$ depends on the sign of $c_5$; for $c_5 = +1$, the cross sections monotonically increase with $\Lambda$ getting lower, while for $c_5 = -1$, there is cancellation between the tree and loop contributions and the cross sections can be below the neutrino floor for certain values of $\Lambda$.

The SD DM-nucleon scattering is induced by the exchange of the $Z$ boson. The DM can couple to the $Z$ boson through the following dimension-six operator\footnote{We also have the dimension-six operator of the form 
\begin{align}
  \mathcal{L}_6' &= \frac{c_6'}{\Lambda^2} 
  \left( H^\dagger \tau_a i {D}_\mu H - i (D_\mu H)^\dagger \tau_a H \right)
   \sum_{m,n} \chi_m^\dagger (T_a)_{mn } \bar{\sigma}^\mu \chi_n ~,
   \label{eq:l6pr}
\end{align} 
but this does not induce the $Z$-boson coupling to the DM component $\chi_0$. 
} 
\begin{align}
  \mathcal{L}_6 &= \frac{c_6}{\Lambda^2} H^\dagger i \overleftrightarrow{D}_\mu H \sum_{m = -j}^j \chi_m^\dagger \bar{\sigma}^\mu \chi_m ~,
  \label{eq:l6}
\end{align} 
where $H^\dagger i \overleftrightarrow{D}_\mu H \equiv H^\dagger i D_\mu H - i (D_\mu H)^\dagger H$. This effective operator induces the DM-$Z$ boson coupling after the Higgs field acquires a VEV. This then generates the SD DM-quark couplings 
\begin{align}
  d_q^{(6)} = \frac{c_6}{2 \Lambda^2} T_q^3 ~,
  \label{eq:dq6}
\end{align}
with $T_q^3 = 1/2$ for $q = u$ and $-1/2$ for $q = d,s$. If the DM has direct couplings to quarks through heavy states, we also have the effective operators of the form 
\begin{equation}
  \mathcal{L}_6^{\prime \prime} = \sum_q \frac{c_q}{\Lambda^2} q^{\dagger} \bar{\sigma}_\mu q \sum_{m = -j}^j \chi_m^\dagger \bar{\sigma}^\mu \chi_m ~,
  \label{eq:l6pr2}
\end{equation}
whose coefficients are the same order as $d_q^{(6)}$ in Eq.~\eqref{eq:dq6}. To avoid unnecessary complications, we just set $c_q = 0$ in the following analysis.

In addition, we have the electroweak loop contribution~\cite{Hisano:2011cs}
\begin{align}
  d_q^{(\mathrm{EW})} = \frac{n^2 -1}{8} \frac{\alpha_2^2}{m_W^2} g_{\mathrm{AV}} \biggl( \frac{m_W^2}{m_{\mathrm{DM}}^2}\biggr) ~,
\end{align}
where 
\begin{equation}
  g_{\mathrm{AV}} (x) = \frac{1}{24 b_x} \sqrt{ x} (8 -x - x^2) \tan^{-1} \biggl(\frac{2 b_x}{\sqrt{x}}\biggr) - \frac{1}{24} x \left[ 2-(3+x)\ln x \right] ~,
\end{equation}
with $b_x \equiv \sqrt{1 -x/4}$. The total SD DM-quark couplings are then given by $d_q = d_q^{(6)} + d_q^{(\mathrm{EW})} $. These couplings give rise to the SD DM-nucleon couplings 
\begin{equation}
  a_N  = \sum_{q = u,d,s} d_q \Delta q_N ~,
  \label{eq:an}
\end{equation}
where $\Delta q_N$ are the spin fractions: $2 s_\mu \Delta q_N \equiv \langle N | \bar{q} \gamma_\mu \gamma_5 q | N \rangle$, with $s_\mu$ the spin four-vector of the nucleon. For $\Delta q_N$, we use the values obtained by QCD lattice simulations: $\Delta u_p = 0.862(17)$, $\Delta u_n = - 0.424(16)$, $\Delta d_p = - 0.424(16)$, $\Delta d_n = 0.862(17)$, $\Delta s_p = \Delta s_n = -0.0458(73)$~\cite{Alexandrou:2019brg}. The SD scattering cross section is then obtained as 
\begin{equation}
  \sigma_{\mathrm{SD}}^{(N)} = \frac{12}{\pi} \biggl(\frac{m_N m_{\mathrm{DM}}}{m_N + m_{\mathrm{DM}}}\biggr)^2 a_N^2 ~. 
  \label{eq:sigsdN}
\end{equation}
This formula is valid in the limit of zero momentum transfer. This is a fairly good approximation for the evaluation of the DM direct detection rate, but may be modified by a factor of $\sim 10$ for the scattering in NSs~\cite{Bell:2020obw, Anzuini:2021lnv}.

\begin{figure}
  \centering
  \subcaptionbox{\label{fig:sigsd3}
  $n = 3$
  }
  {\includegraphics[width=0.48\textwidth]{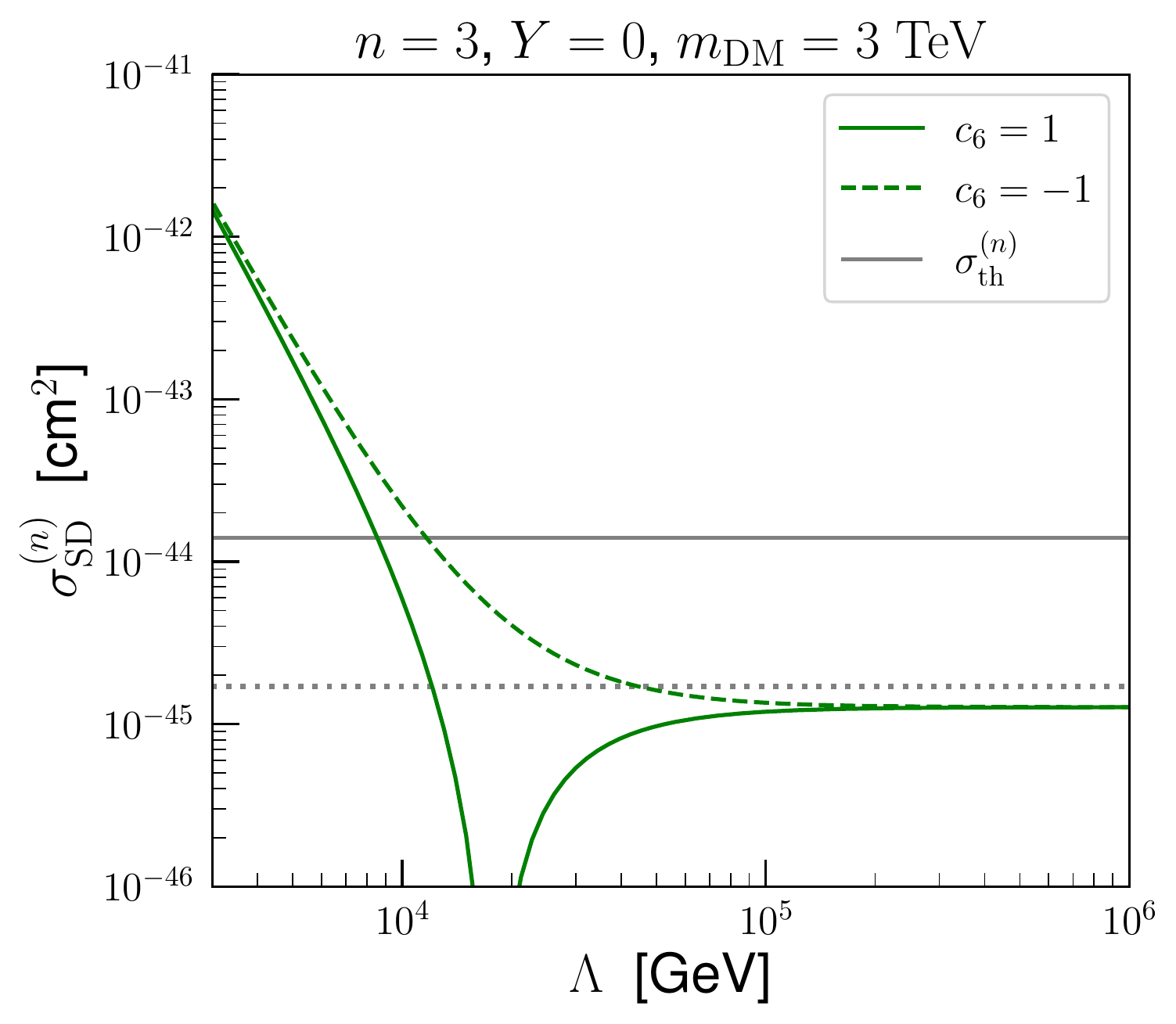}}
  \subcaptionbox{\label{fig:sigsd5}
  $n = 5$
  }
  { 
  \includegraphics[width=0.48\textwidth]{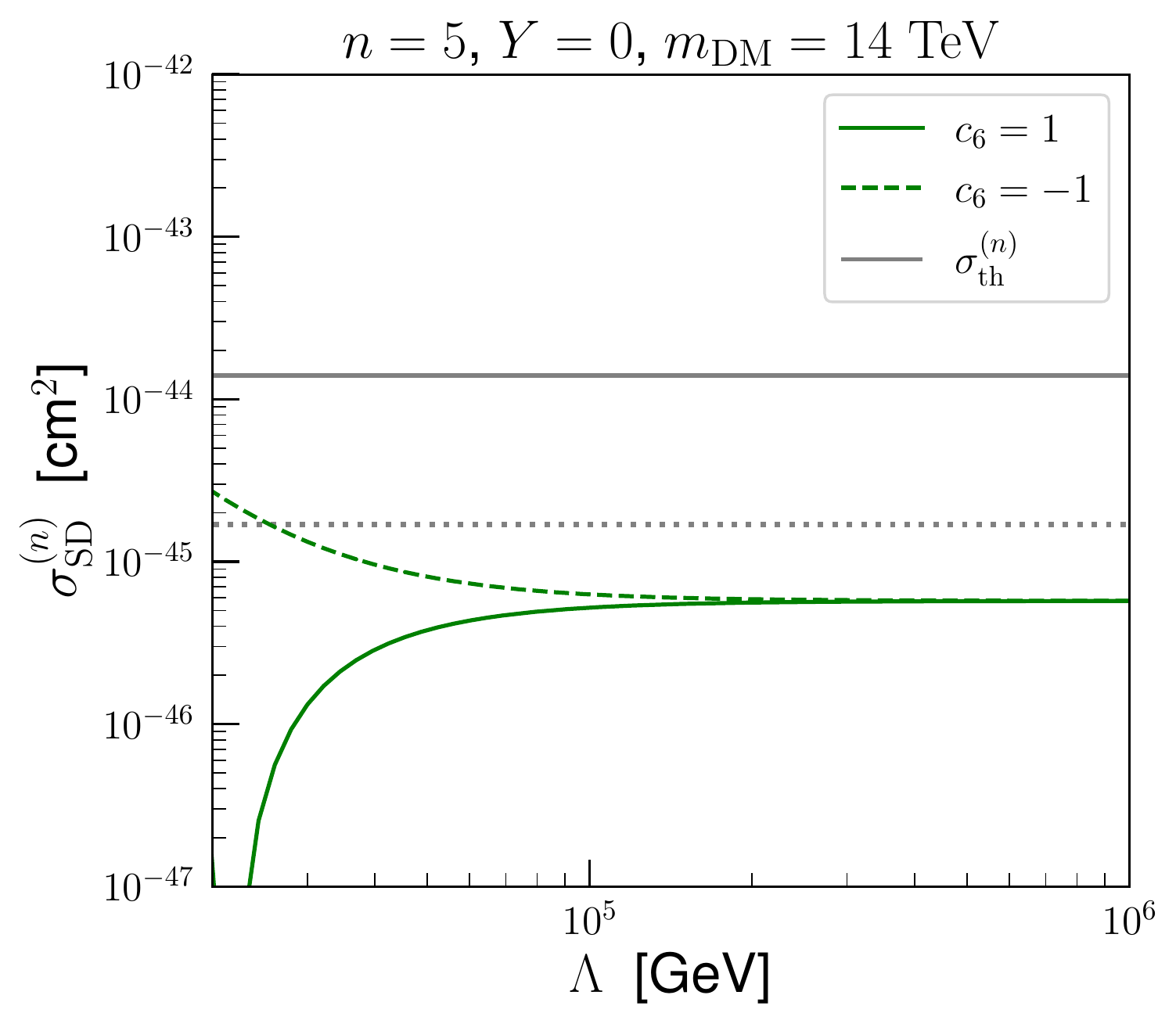}}
  \caption{
    The SD DM-neutron elastic scattering cross section {for $Y=0$,} as a function of the cut-off scale $\Lambda$ for (a) $n = 3$ and $m_{\rm DM} = 3$~TeV and (b) $n=5$ and $m_{\rm DM} = 14$~TeV. The green solid and dashed lines correspond to the cases with $c_6 = +1$ and $-1$, respectively. The solid and dotted horizontal gray lines indicate the threshold cross section $\sigma^{(n)}_{\mathrm{th}}$ estimated in Ref.~\cite{Anzuini:2021lnv} and Ref.~\cite{Bell:2020jou}, respectively. 
  }
  \label{fig:sdcrossy0}
\end{figure}

In Fig.~\ref{fig:sdcrossy0}, we show the SD DM-neutron elastic scattering cross section as a function of the cut-off scale $\Lambda$ for $n = 3$ and $m_{\rm DM} = 3$~TeV (Fig.~\ref{fig:sigsd3}) and $n=5$ and $m_{\rm DM} = 14$~TeV (Fig.~\ref{fig:sigsd5}). The green solid and dashed lines correspond to the cases with $c_6 = +1$ and $-1$, respectively. We also show the threshold cross section $\sigma^{(n)}_{\mathrm{th}} \simeq 1.7 \times 10^{-45}~\mathrm{cm}^2$ obtained in Ref.~\cite{Bell:2020jou} in the horizontal dotted line and  $\sigma^{(n)}_{\mathrm{th}} \simeq 1.4 \times 10^{-44}~\mathrm{cm}^2$ obtained in Ref.~\cite{Anzuini:2021lnv} in the horizontal solid line. In the latter, the effect of large momentum transfer and effective nucleon mass in NSs is taken into account. The difference between these two values may be regarded as theoretical uncertainty in the estimation of $\sigma_{\mathrm{th}}^{(n)}$. The current limits from the DM direct detection experiments (see, \textit{e.g.}, Refs.~\cite{PICO:2019vsc, XENON:2019rxp}), as well as that imposed by the IceCube experiment~\cite{IceCube:2016dgk}, are above the range of the cross sections shown in Fig.~\ref{fig:sdcrossy0}. As we see in these figures, the SD cross sections also converge into certain values determined by the electroweak loop contribution\footnote{These values are larger than those given in Ref.~\cite{Hisano:2011cs} by an almost order of magnitude---the difference originates from $\Delta q_N$. Since $d_q^{(\mathrm{EW})}$ is identical for $q = u, d, s$, its contribution to $a_N$ is simply given by $d_q^{(\mathrm{EW})} \sum_{q = u,d,s} \Delta q_N$. For the spin fractions used in this paper, $\sum_{q = u,d,s} \Delta q_N \simeq 0.39$, while for those used in Ref.~\cite{Hisano:2011cs}, $\sum_{q = u,d,s} \Delta q_N \simeq 0.13$. The former is larger than the latter by a factor of $\sim 3$, which leads to roughly an order-of-magnitude difference in the scattering cross section. } in the limit of large $\Lambda$. For $n = 3$, the SD cross section can be larger than the threshold cross section for $\Lambda \lesssim 10$~TeV, while for $n=5$, it is generically predicted to be lower than the threshold value.

\subsubsection{Inelastic scattering}
\label{eq:inelasticy0}

The inelastic scattering processes, such as $\chi^0 + n \to \chi^- + p$, occur in NSs if $\Delta M_\pm < \Delta M_{\mathrm{max}}$.\footnote{As discussed in footnote \ref{ft:mass_splitting}, the threshold $\Delta M_{\rm  max}$ suffers from theoretical uncertainties in the modeling of dense matter.} This process is induced by the exchange of the $W$ boson; the interaction of the DM with the $W$ boson is given by 
\begin{equation}
  \mathcal{L}_{\mathrm{int}} = \frac{g_2}{2\sqrt{2}} \sqrt{n^2-1} \, \overline{\psi^-} \gamma^\mu \psi^0\, W_\mu^-  + \mathrm{h.c.} ~,
\end{equation}
where we use the four-component notation 
\begin{equation}
  \psi^0 \equiv 
  \begin{pmatrix}
    \chi_0 \\ \chi_0^\dagger 
  \end{pmatrix}
  ~, \qquad 
  \psi^- \equiv 
  \begin{pmatrix}
    \chi_{-1} \\ \ - \chi_{1}^\dagger 
  \end{pmatrix}
  ~.
\end{equation}
This interaction induces the following four fermion operator 
\begin{align}
  \mathcal{L}_{\mathrm{CC}} = - \frac{g_2^2 V_{ud}}{4 m_W^2} \sqrt{n^2 -1} \, \overline{\psi^-} \gamma^\mu \psi^0\, \bar{u} \gamma_\mu P_L d + \mathrm{h.c.} ~,
  \label{eq:lcc}
\end{align}
where $P_L \equiv (1-\gamma_5)/2$ and $V_{ud}$ is the CKM matrix element. 

The cross sections of the inelastic scattering processes induced by the  effective interaction in Eq.~\eqref{eq:lcc} turn out to be much larger than the threshold cross section unless the charged-neutral mass splitting $\Delta M_\pm$ is very close to the threshold value $\Delta M_{\mathrm{max}}$. To see this, it is sufficient to calculate the cross section for $\Delta M_\pm \to 0$ in the non-relativistic approximation. We have 
\begin{align}
  \sigma(\chi^0 + n \to \chi^- + p) & \simeq 
  \frac{G_F^2 V_{ud}^2}{2\pi} (n^2-1) \biggl(\frac{m_N m_{\mathrm{DM}}}{m_N + m_{\mathrm{DM}}}\biggr)^2 \\[3pt]
  & \simeq 7 (n^2-1) \times 10^{-39}~\mathrm{cm}^2 ~,
\end{align}
where $G_F$ is the Fermi constant. This is larger than the threshold cross section by orders of magnitude, implying that the DM can efficiently be captured by NSs if $\Delta M_{\pm} < \Delta M_{\mathrm{max}}$.

\subsubsection{Implications for DM heating in NSs}

As discussed in Sec.~\ref{sec:yzero}, the neutral-charged mass splitting in Eq.~\eqref{eq:delmplmiy0}, $\Delta M_{\pm}$, is predicted to be smaller than the threshold mass difference in Eq.~\eqref{eq:delmmax}, $\Delta M_{\mathrm{max}}$. We find $\Delta M_{\pm} \simeq \Delta M_{\mathrm{max}}/2$, with little dependence on the cutoff scale $\Lambda$. Consequently, the DM-nucleon inelastic scattering can occur in NSs for $Y = 0$. This scattering cross section is much larger than the threshold cross section, as we see {have seen} in Sec.~\ref{eq:inelasticy0}. In addition, for $n = 3$, the SD elastic scattering cross section is also larger than the threshold cross section for $\Lambda \lesssim 10$~TeV. We therefore conclude that the DM can efficiently be captured by NSs for $Y = 0$, and the NS surface temperature is heated up to the maximum value, $\simeq \text{a few} \times 10^3$~K, in old NSs because of the DM heating effect.

\subsection{$Y \neq 0$}
\label{sec:ynonzerodn}

\subsubsection{Elastic scattering}

We now consider the $Y \neq 0$ case. The DM-nucleon SI elastic scattering is induced by the dimension-five operators in Eq.~\eqref{eq:l5ynon0}. These operators generate the DM-nucleon SI effective coupling via the Higgs-boson exchange. For $Y \geq 1$, we have 
\begin{equation}
  f_N^{(5)} =  \frac{m_N}{2 \Lambda m_h^2} \biggl(c_5 + \frac{Y}{2} c_5' \biggr) \biggl[\sum_{q = u,d,s} f_{T_q}^{(N)} + 3 \times \frac{2}{27} f_{T_G}^{(N)} \biggr] ~.
\end{equation}
For $Y = 1/2$, on the other hand, the operators in Eq.~\eqref{eq:effc} are also of dimension-five, and therefore their contribution is comparable to that of Eq.~\eqref{eq:l5ynon0}. In this case, we have 
\begin{equation}
  f_N^{(5)} =  \frac{m_N}{2 \Lambda m_h^2} \biggl(c_5 + \frac{1}{4} c_5' - \frac{|c_s + c_s'|}{2}\biggr) \biggl[\sum_{q = u,d,s} f_{T_q}^{(N)} + 3 \times \frac{2}{27} f_{T_G}^{(N)} \biggr] ~.
  \label{eq:fn5y12}
\end{equation}
Similarly to the case of $Y = 0$, the DM-nucleon scattering is induced also by the electroweak gauge-boson loop diagrams. In the present case, the $Z$ boson also contributes to the process, and instead of Eq.~\eqref{eq:fnewy0}, we have 
\begin{equation}
  f_N^{(\rm EW)} =  (n^2 - 4Y^2 -1) f_N^W + Y^2 f^Z_N ~,
  \label{eq:fnewynon0}
\end{equation}
where in the limit of $m_{\mathrm{ DM}} \gg m_W, m_Z$~\cite{Hisano:2015rsa}, 
\begin{equation}
  f_p^Z \simeq - 1.9 \times 10^{-10} ~{\rm GeV}^{-2} ~, \qquad 
  f_n^Z \simeq - 1.8 \times 10^{-10} ~{\rm GeV}^{-2} ~.
\end{equation}
The effective DM-nucleon coupling is then given by $f_N = f_N^{(5)} + f_N^{(\mathrm{EW})}$. 

\begin{figure}
  \centering
  \subcaptionbox{\label{fig:sigsi2}
  $n = 2, Y = 1/2$
  }
  {\includegraphics[width=0.48\textwidth]{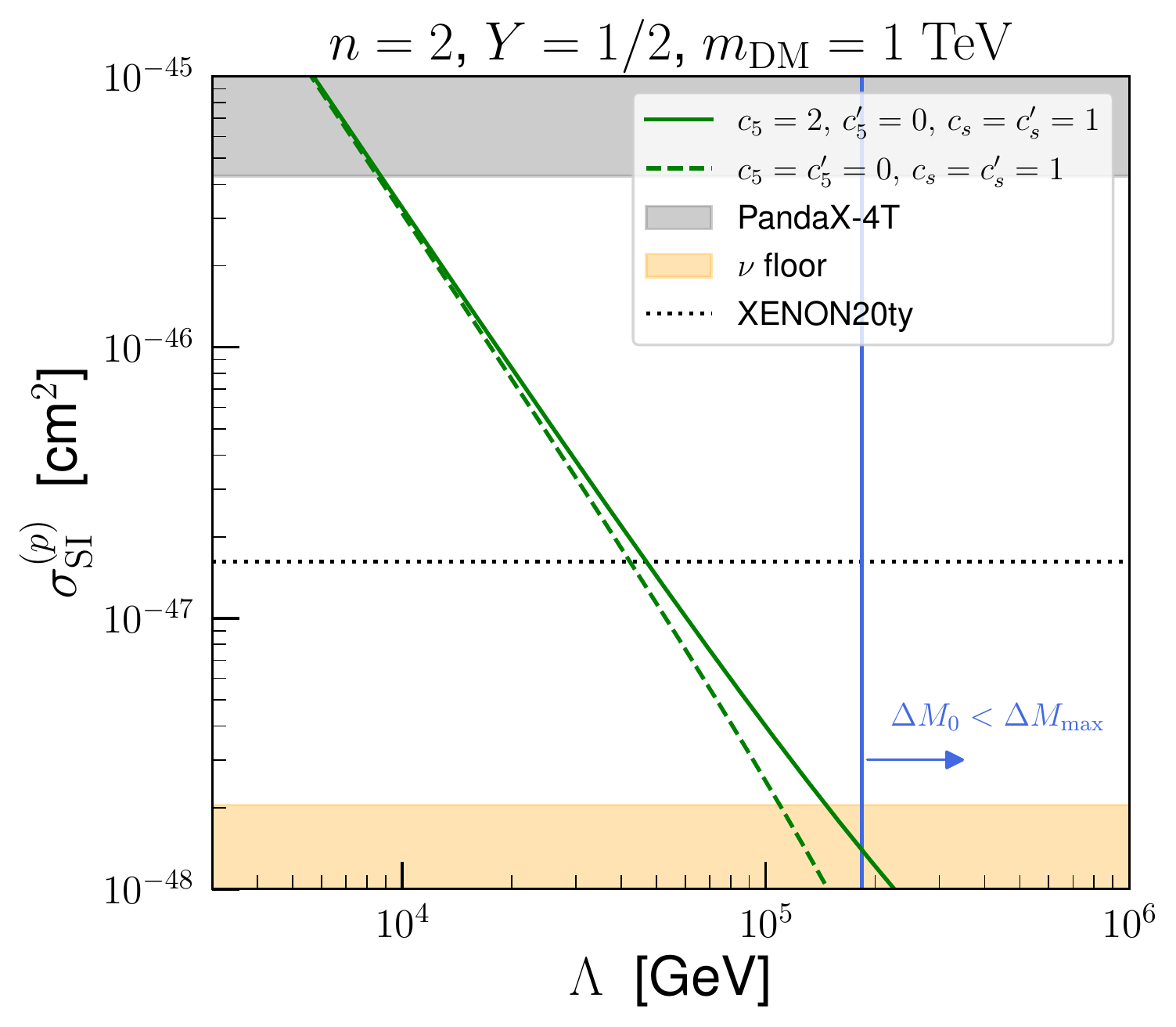}}
  \subcaptionbox{\label{fig:sigsi3_1}
  $n = 3, Y = 1$
  }
  {\includegraphics[width=0.48\textwidth]{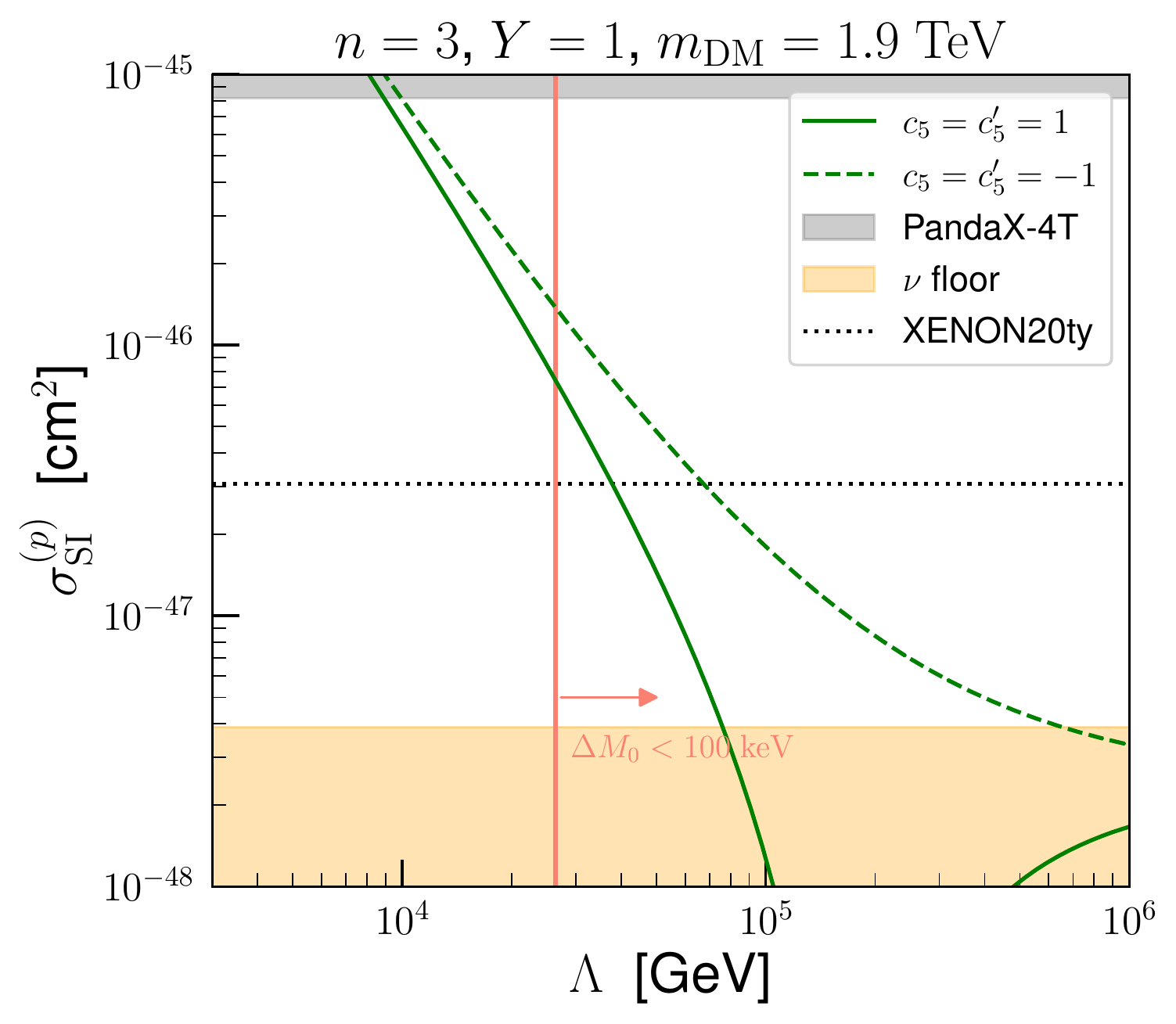}}
  \subcaptionbox{\label{fig:sigsi4}
  $n = 4, Y = 3/2$
  }
  {\includegraphics[width=0.48\textwidth]{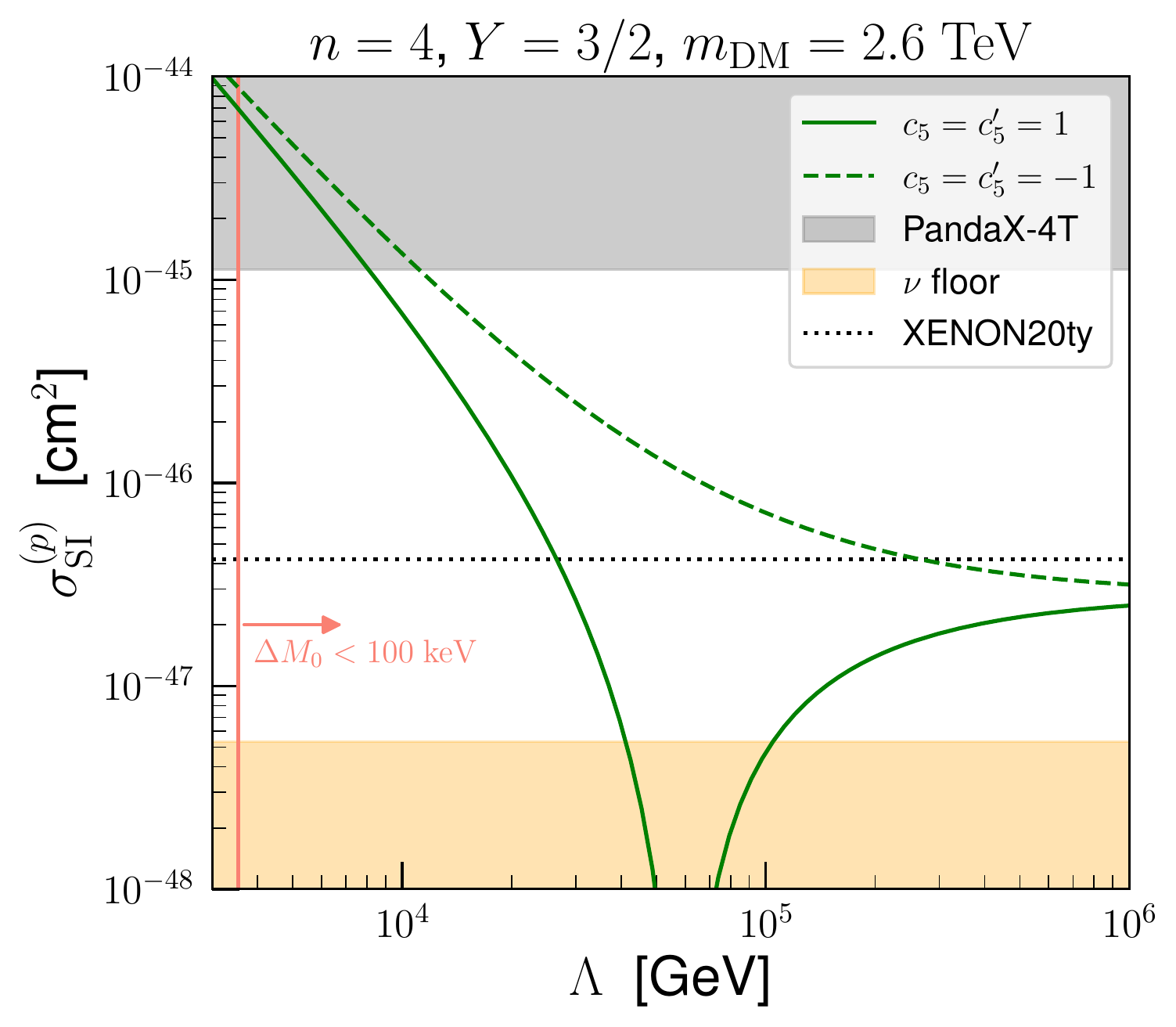}}
  \caption{
    The SI DM-proton elastic scattering cross section as a function of the cut-off scale $\Lambda$ for (a) $n = 2, Y=1/2$ and $m_{\rm DM} = 1$~TeV, (b) $n=3, Y =1$ and $m_{\rm DM} = 1.9$~TeV, (c) $n=4, Y = 3/2$ and $m_{\mathrm{DM}} = 2.6~\mathrm{TeV}$. The gray shaded area is excluded by the PandaX-4T experiment~\cite{PandaX-4T:2021bab}. The horizontal dotted line represents the expected sensitivity of the XENONnT experiment with a 20 ton-year exposure~\cite{XENON:2020kmp}, and the orange area corresponds to the neutrino floor~\cite{Billard:2021uyg}. {The blue and the orange vertical lines correspond to the threshold mass differences that permit DM inelastic scatterings in NSs and direct detection experiments, respectively, for $c_s = c_s' = 1$. }
  }
  \label{fig:sicrossyno0}
  \end{figure}

In Fig.~\ref{fig:sicrossyno0}, we show the SI DM-proton elastic scattering cross sections as functions of the cut-off scale $\Lambda$ for $n = 2, Y=1/2$ and $m_{\rm DM} = 1$~TeV (Fig.~\ref{fig:sigsi2}), $n=3, Y =1$ and $m_{\rm DM} = 1.9$~TeV (Fig.~\ref{fig:sigsi3_1}), and $n=4, Y = 3/2$ and $m_{\mathrm{DM}} = 2.6~\mathrm{TeV}$ (Fig.~\ref{fig:sigsi4}). In Fig.~\ref{fig:sigsi2}, the green solid (dashed) line corresponds to $c_5 = 2, \, c_5' = 0, \, c_s = c_s' = +1$ ($c_5 =  c_5' = 0, \, c_s = c_s' = +1$);\footnote{We set these parameters such that the combination in the parenthesis in Eq.~\eqref{eq:fn5y12} equals to $\pm 1$. } in Fig.~\ref{fig:sigsi3_1} and Fig.~\ref{fig:sigsi4}, the solid and dashed lines are for $c_5 = c_5' = +1$ and $-1$, respectively. The gray shaded area is excluded by the PandaX-4T experiment~\cite{PandaX-4T:2021bab}. The horizontal dotted line represents the expected sensitivity of the XENONnT experiment with a 20 ton-year exposure~\cite{XENON:2020kmp}, and the orange area corresponds to the neutrino floor~\cite{Billard:2021uyg}.  We see that the PandaX-4T limit has already excluded lower regions of $\Lambda$. This sets $\Lambda \gtrsim 9$~TeV, 10~TeV, and 8~TeV for the doublet, triplet, and quartet cases, respectively. 

As we have seen in Sec.~\ref{sec:ynonzero}, DM direct detection experiments also impose upper limits on $\Lambda$ which sets $\Lambda \lesssim \text{a few} \times 10^{4}$~GeV and a few TeV for the triplet and quartet DM, respectively, in order to suppress the inelastic scattering. This excludes the region to the right of the orange vertical lines in Fig.~\ref{fig:sicrossyno0}. As a result, the quartet DM has already been ruled out---we thus do not consider this case in the following discussion. There is still a small allowed range of $\Lambda$ {near $10$\,TeV} for the triplet case, which will fully be explored in the XENONnT experiment.

Next, we consider the SD DM-nucleon scattering, which is again induced by the $Z$-boson exchange. The DM-$Z$ boson coupling is given by 
\begin{align}
  \mathcal{L}_{\chi Z} & = - g_Z Y \left[ \left|\left( U \right)_{11} \right|^2 - \left|\left( U \right)_{12} \right|^2 \right] \left( \chi^0_1 \right)^\dagger \bar{\sigma}^\mu \chi^0_1 \, Z_\mu \nonumber \\ 
  & \simeq \frac{g_Z Y}{2 M} \biggl(\frac{v^{4Y}}{2^{2Y}\Lambda^{(4Y-1)}}\biggr) \left( c_s - c_s' \right) \mathrm{sgn} \left( c_s + c_s' \right) 
  \left( \chi^0_1 \right)^\dagger \bar{\sigma}^\mu \chi^0_1 \, Z_\mu ~.
\end{align} 
This gives the dominant contribution for $Y = 1/2$. For $Y = 1$, on the other hand, dimension-six operators similar to Eq.~\eqref{eq:l6} give the dominant contribution:\footnote{In addition to these terms, we have terms similar to Eq.~\eqref{eq:l6pr} and Eq.~\eqref{eq:l6pr2}, whose contribution to $d_q$ is the same order as those of Eq.~\eqref{eq:l6ynon0}. We suppress the coefficients of these operators in the following analysis for simplicity.  } 
\begin{align}
  \mathcal{L}_6 &= \frac{c_6}{\Lambda^2} H^\dagger i \overleftrightarrow{D}_\mu H \sum_{m = -j}^j \chi_m^\dagger \bar{\sigma}^\mu \chi_m
  +\frac{c_6'}{\Lambda^2} H^\dagger i \overleftrightarrow{D}_\mu H \sum_{m = -j}^j \eta_m^\dagger \bar{\sigma}^\mu \eta_m 
  ~.
  \label{eq:l6ynon0}
\end{align}
These interactions induce the SD DM-quark couplings 
\begin{align}
  d_q^{(\mathrm{tree})} &= - \frac{1}{8 M \Lambda} (c_s -c_s') \,\mathrm{sgn} (c_s + c_s') \,T_q^3 ~,
  \label{eq:dqtree}
\end{align}
for $Y = 1/2$ and 
\begin{align}
  d_q^{(\mathrm{tree})} &= \frac{c_6 + c_6'}{4 \Lambda^2} T_q^3 ~, 
  \label{eq:dqtreey1}
\end{align}
for $Y = 1$. We also have the electroweak loop contributions for both $Y = 1/2$ and $1$: 
\begin{align}
  d_q^{(\mathrm{EW})} = \frac{n^2 -(4Y^2 +1)}{8} \frac{\alpha_2^2}{m_W^2} g_{\mathrm{AV}} \biggl( \frac{m_W^2}{m_{\mathrm{DM}}^2}\biggr) 
  +\frac{2 Y^2 \left[ \left( a_q^V \right)^2 + \left( a_q^A \right)^2 \right] }{\cos^4 \theta_W} \frac{\alpha_2^2}{m_Z^2} g_{\mathrm{AV}} \biggl( \frac{m_Z^2}{m_{\mathrm{DM}}^2}\biggr) 
  ~,
\end{align}
where 
\begin{equation}
  a_q^V = \frac{1}{2} T^3_q - Q_q \sin^2 \theta_W ~, \qquad a_q^A = -\frac{1}{2} T^3_q ~.
\end{equation}
The SD scattering cross section is then computed by using Eq.~\eqref{eq:an} and \eqref{eq:sigsdN}, with $d_q = d_q^{(\mathrm{tree})} + d_q^{(\mathrm{EW})}$.

\begin{figure}
  \centering
  \subcaptionbox{\label{fig:sigsd2n}
  $n = 2,\, Y = 1/2$
  }
  {\includegraphics[width=0.48\textwidth]{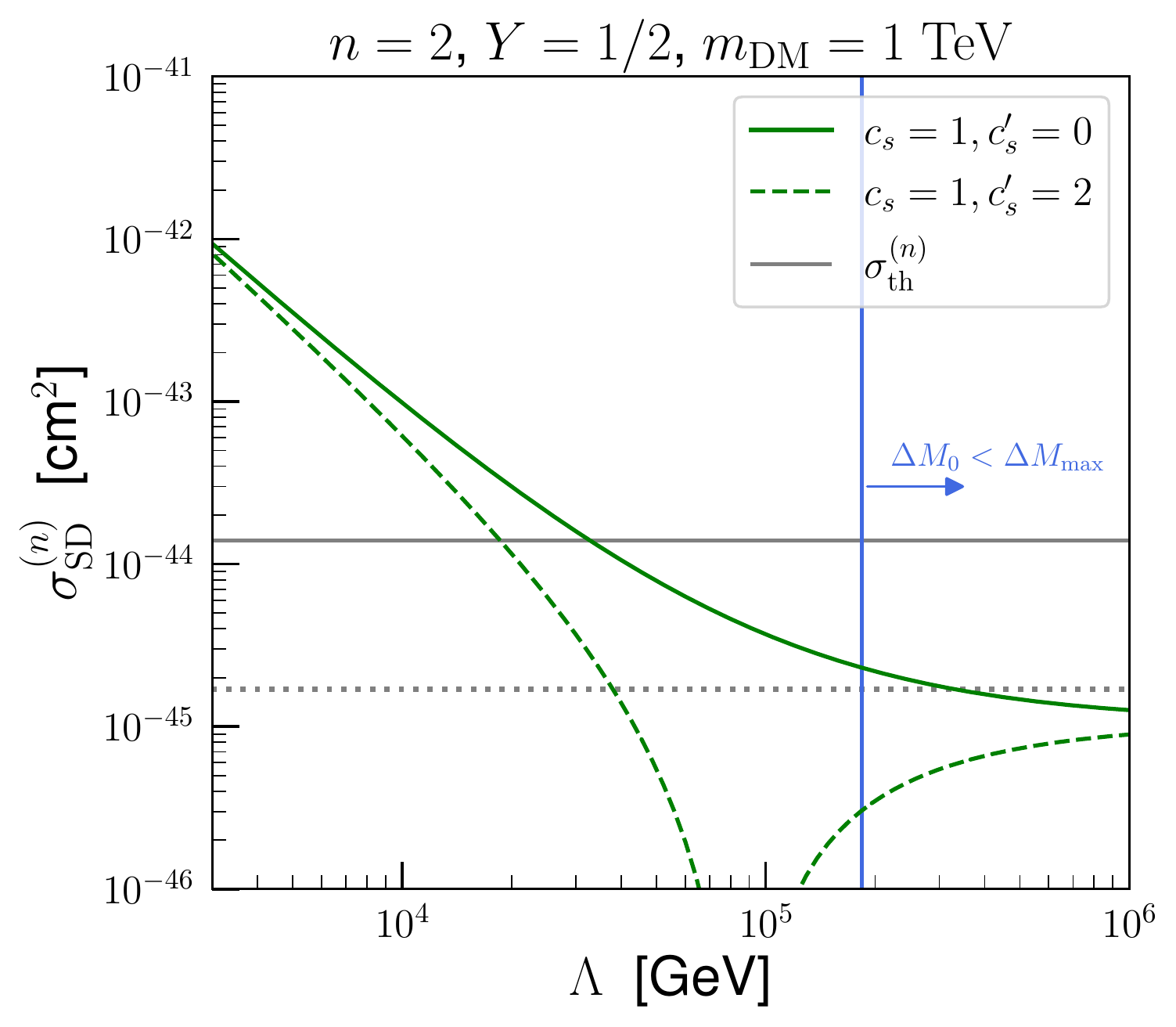}}
  \subcaptionbox{\label{fig:sigsdy1n}
  $n = 3,\, Y=1$
  }
  { 
  \includegraphics[width=0.48\textwidth]{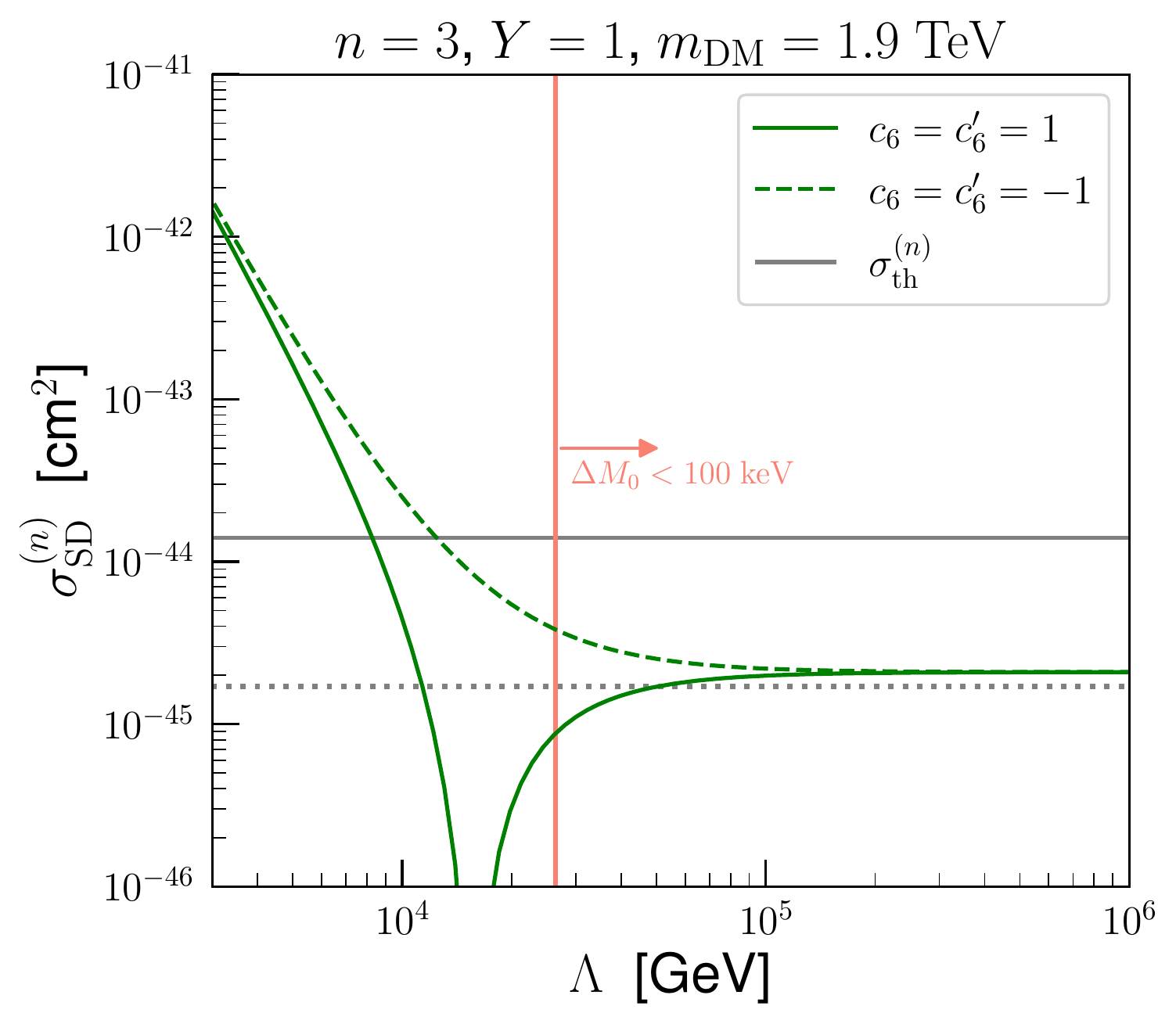}}
  \caption{
    The SD DM-neutron elastic scattering cross section as a function of the cut-off scale $\Lambda$ for (a) $n = 2$, $Y=1/2$, and $m_{\rm DM} = 1$~TeV with $c_s = 1,\, c_s' =0$ (solid) and $c_s = 1, \, c_s' = 2$ (dashed); (b) $n=3$, $Y=1$, and $m_{\rm DM} = 1.9$~TeV with $c_6 = c_6' =1$ (solid) and $c_6 = c_6' = -1$ (dashed). The solid and dotted horizontal gray lines indicate the threshold cross section $\sigma^{(n)}_{\mathrm{th}}$ estimated in Ref.~\cite{Anzuini:2021lnv} and Ref.~\cite{Bell:2020jou}, respectively. {The blue and the orange vertical lines correspond to the threshold mass differences that permit DM inelastic scatterings in NSs and direct detection experiments, respectively, for $c_s = c_s' = 1$.}
  }
  \label{fig:sdcrossynon0}
\end{figure}

In Fig.~\ref{fig:sdcrossynon0}, we plot the SD DM-neutron elastic scattering cross section as a function of the cut-off scale $\Lambda$ for (a) $n = 2$, $Y=1/2$, and $m_{\rm DM} = 1$~TeV with $c_s = 1,\, c_s' =0$ (solid) and $c_s = 1, \, c_s' = 2$ (dashed);\footnote{These two choices correspond to $(c_s -c_s')\, \mathrm{sgn} (c_s + c_s') = \pm 1$ in Eq.~\eqref{eq:dqtree}. } (b) $n=3$, $Y=1$, and $m_{\rm DM} = 1.9$~TeV with $c_6 = c_6' =1$ (solid) and $c_6 = c_6' = -1$ (dashed). The solid and dotted horizontal gray lines indicate the threshold cross section $\sigma^{(n)}_{\mathrm{th}}$ estimated in Ref.~\cite{Anzuini:2021lnv} and Ref.~\cite{Bell:2020jou}, respectively. We see that the SD scattering cross section exceeds the threshold cross section if $\Lambda \lesssim 20$~TeV ($8$~TeV) in the case of $n = 2,\, Y = 1/2$ ($n = 3,\, Y=1$). It is also found that even in the large $\Lambda$ limit, the SD cross sections remain sizable because of the electroweak-loop corrections. Recall that the capture rate is suppressed by a factor of $\sigma^{(n)}_{\mathrm{SD}}/\sigma^{(n)}_{\mathrm{th}}$ and thus the suppression in the surface temperature goes as $(\sigma^{(n)}_{\mathrm{SD}}/\sigma^{(n)}_{\mathrm{th}})^{1/4}$ (see Eqs.~(\ref{eq:lh}--\ref{eq:lhlgameq})). As a consequence, there is little decrease in the surface temperature even if $\Lambda$ is very large and may still be detectable in future observations.

\subsubsection{Inelastic scattering}

In the case of $Y \neq 0$, not only the charged-current processes discussed in Sec.~\ref{eq:inelasticy0} but also the neutral-current processes $\chi^0_1 + N \to \chi^0_2 + N$ can occur via the $Z$-boson exchange. The relevant interaction in this case is 
\begin{align}
  \mathcal{L}_{\mathrm{int}} = i g_Z Y (\chi_2^0)^\dagger \bar{\sigma}^\mu \chi_1^0 Z_\mu + \mathrm{h.c.} ~.
\end{align}
Using this interaction, we estimate the inelastic scattering cross section in the non-relativistic limit for $\Delta M_0 \to 0$ as 
\begin{align}
  \sigma ( \chi^0_1 + N \to \chi^0_2 + N) &\simeq \frac{G_F^2 Y^2}{2\pi} c_N^2  \biggl(\frac{m_N m_{\mathrm{DM}}}{m_N + m_{\mathrm{DM}}}\biggr)^2 \\[3pt]
  & \simeq 7 Y^2 c_N^2 \times 10^{-39}~\mathrm{cm}^2 ~,
\end{align}
where $c_n = 1$ and $c_p = 1- 4\sin^2\theta_W$. Again, this scattering cross section is much larger than the threshold correction. Taking account of the discussion in Sec.~\ref{eq:inelasticy0}, we conclude that the $Y\neq 0$ DM candidates are efficiently captured by NSs if $\Delta M_0$ or $\Delta M_\pm$ is smaller than $\Delta M_{\mathrm{max}}$. {This threshold is indicated by the blue vertical lines in Fig.~\ref{fig:sicrossyno0} and \ref{fig:sdcrossynon0}.}

\subsubsection{Implications for DM heating in NSs}

For $n = 3$ and $Y = 1$, the PandaX-4T bound imposes a lower limit on the cut-off scale $\Lambda$ as $\Lambda \gtrsim 10$~TeV, as we have seen in Fig.~\ref{fig:sigsi3_1}. Figure~\ref{fig:delm0} then shows that in this allowed region the mass difference between the neutral components is always smaller than $\Delta M_{\mathrm{max}}$, and thus the inelastic scattering $\chi^0_1 + n \to \chi^0_2 + n$ can occur. This means that the DM heating effect operates in NSs with the maximal capture rate for this DM candidate. 

For $n = 2$ and $Y = 1/2$, on the other hand, the inelastic scattering can occur for $\Lambda \gtrsim 200$~TeV, as shown in Fig.~\ref{fig:delm0}, while the SD elastic scattering cross section exceeds the threshold cross section for $\Lambda \lesssim 20$~TeV, as seen in Fig.~\ref{fig:sigsd2n}. For $20~\mathrm{TeV} \lesssim \Lambda \lesssim 200~\mathrm{TeV}$, the capture rate is suppressed by a factor of $\sigma^{(n)}_{\mathrm{SD}}/\sigma^{(n)}_{\mathrm{th}}$. We, however, note that since the resultant decrease in temperature is only by a factor of $(\sigma^{(n)}_{\mathrm{SD}}/\sigma^{(n)}_{\mathrm{th}})^{1/4}$, except for the cancellation region seen for $c_s = 1, \, c_s' = 2$ in Fig.~\ref{fig:sigsd2n}, the NS surface temperature is still $\gtrsim 10^3$~K, and thus can be observed with future infrared telescopes. 

We also note that for the doublet DM the XENONnT experiment can probe up to $\Lambda \sim 40~\mathrm{TeV}$
{(see Fig.~\ref{fig:sigsi2})}. Moreover, for $\Lambda \lesssim 100$~TeV, the SI DM-nucleon elastic scattering cross section is above the neutrino floor, and thus can potentially be tested in future DM direct searches. This shows that the DM search by means of the NS temperature observation and DM direct detection experiments play complementary roles for the doublet case.

\section{Conclusion and discussion}
\label{sec:conclusion}

\begin{figure}
  \centering
  \includegraphics[width=1\textwidth]{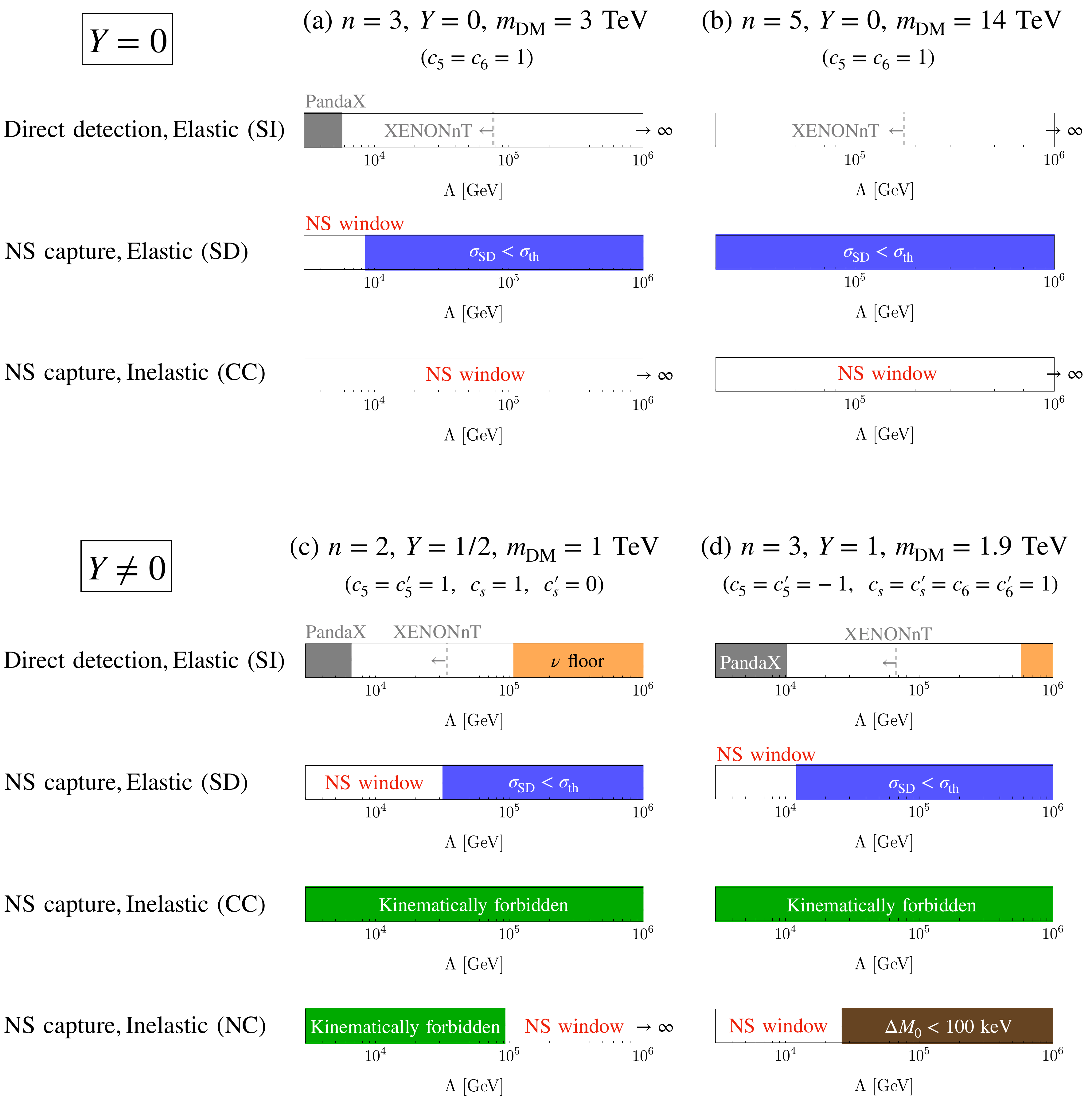}
  \caption{
    {The current constraints and future prospects of DM direct detection experiments, as well as the potential reach of NS temperature observation (\textit{NS window}), are shown in terms of the cut-off scale $\Lambda$ for 
    (a) $n = 3$, $Y=0$, and $m_{\rm DM} = 3~\mathrm{TeV}$ with $c_5 = c_6 = 1$;
    (b) $n = 5$, $Y=0$, and $m_{\rm DM} = 14~\mathrm{TeV}$ with $c_5 = c_6 = 1$;
    (c) $n = 2$, $Y=1/2$, and $m_{\rm DM} = 1~\mathrm{TeV}$ with $c_5 = c_5' = 1, \, c_s = 1,\, c_s' =0$;
    (d) $n = 3$, $Y=1$, and $m_{\rm DM} = 1.9~\mathrm{TeV}$ with $c_5 = c_5' = -1, \, c_s = c_s' = c_6 = c_6' = 1$. 
    The gray and orange bands represent the current DM direct detection bounds and the neutrino floor, respectively. The gray dashed lines indicate the sensitivity of XENONnT. We have $\sigma_{\rm  SD}  <  \sigma_{\rm  th}  \simeq  1.4  \times 10^{-44}~\mathrm{cm}^2$ in the blue region, where the predicted NS temperature is suppressed by a factor of $(\sigma_{\mathrm{SD}}/\sigma_{\mathrm{th}})^{1/4}$. The inelastic scattering processes are kinematically forbidden in the green regions. The brown region corresponds to $\Delta  M_0  <  100~\mathrm{keV}$ and is excluded by the DM direct detection bound through the inelastic processes.}
  }
  \label{fig:summary}
\end{figure}
%
We have studied electroweak multiplet DM and its scattering cross section with nucleons systematically based on the method of effective field theories. We classify the electroweak DM candidates in terms of their $\mathrm{SU}(2)_L \times \mathrm{U}(1)_Y$ quantum numbers and take account of the effect of UV physics by means of non-renormalizable effective operators suppressed by a heavy mass scale, $\Lambda$. The effect of UV physics is twofold. First, 
as this effect gets stronger, which corresponds to lower $\Lambda$, the DM-nucleon elastic scattering cross section tends to be larger. Second, a lower value of $\Lambda$ results in larger mass splittings among the components of the DM multiplet. If the mass splittings exceed a threshold value, the DM-nucleon inelastic scattering in NSs is suppressed kinematically. We have calculated these quantities as functions of $\Lambda$ to discuss the prospects of testing each model in future DM direct detection experiments and through the temperature observation of old NSs with future infrared telescopes.

Figure~\ref{fig:summary} summarizes our results, which shows the range of $\Lambda$ that can be probed in future DM direct detection experiments and via NS temperature observation (\textit{NS window}), as well as the current limit from the PandaX-4T experiment (gray band). For the $Y = 0$ cases ((a) and (b)), we find that the mass difference between the charge $\pm 1$ and DM components is smaller than the threshold mass difference $\Delta M_{\mathrm{max}}$ and thus the DM-nucleon inelastic scattering in NSs occurs for arbitrary values of $\Lambda$. In addition, low $\Lambda$ regions are within reach of the XENONnT experiment, as indicated by the gray dashed lines, and most of the range of $\Lambda$---even in the limit of $\Lambda \to \infty$---is above the neutrino floor (orange band) thanks to the electroweak-loop contributions. For the triplet DM with hypercharge 1, shown in panel (d), DM direct searches have already narrowed down the allowed range of $\Lambda$; the lower limit is set by PandaX-4T and the upper limit is due to the operation of inelastic scattering, with the mass difference between the neutral components $\lesssim 100$~keV as indicated by the brown band. The remaining allowed region can fully be explored in the XENONnT experiment. Finally, for the doublet DM shown in the panel (c), 
DM capture in NSs is effective through SD elastic scattering for $\Lambda \lesssim 40~\mathrm{TeV}$ and inelastic scattering for $\gtrsim 90~\mathrm{TeV}$ for the parameter choice in Fig.~\ref{fig:summary}. For $40~\mathrm{TeV} \lesssim \Lambda \lesssim 90~\mathrm{TeV}$, the DM capture rate is somewhat suppressed, but the late-time NS temperature is still $\gtrsim 10^3$~K, and thus can be observed with future infrared telescopes. This range can also be probed in future DM direct detection experiments, demonstrating the cooperative roles of direct detection experiments and NS temperature observation.

\section*{Acknowledgments}

We thank Keisuke Yanagi for useful discussions at the early stage of this work. This work is supported in part by JSPS Grant-in-Aid for Scientific Research KAKENHI (No. JP20J12392 [MF]), JSPS Core-to-Core Program (No.JPJSCCA20200002 [MF]), the Grant-in-Aid for Innovative Areas (No.19H05810 [KH], No.19H05802 [KH], No.18H05542 [MF and NN]), Scientific Research B (No.20H01897 [KH and NN]), Young Scientists (No.21K13916 [NN]), and the NSF of China (No.11675086 and 11835005 [JZ]).



\bibliographystyle{utphysmod}
\bibliography{ref}

\providecommand{\href}[2]{#2}\begingroup\raggedright\begin{thebibliography}{100}

\bibitem{Planck:2018vyg}
{\bfseries Planck} Collaboration, {\em {Planck 2018 results. VI. Cosmological
  parameters}}, \href{https://dx.doi.org/10.1051/0004-6361/201833910}{Astron.\
  Astrophys.\  {\bfseries 641} (2020) A6} {\ttfamily
  [\href{https://arxiv.org/abs/1807.06209}{arXiv:1807.06209}]}. [Erratum:
  Astron.Astrophys. 652, C4 (2021)].

\bibitem{Cirelli:2005uq}
M.~Cirelli, N.~Fornengo, and A.~Strumia, {\em {Minimal dark matter}},
  \href{https://dx.doi.org/10.1016/j.nuclphysb.2006.07.012}{Nucl.\  Phys.\
  {\bfseries B753} (2006) 178--194}
{\ttfamily [\href{https://arxiv.org/abs/hep-ph/0512090}{hep-ph/0512090}]}.

\bibitem{Cirelli:2007xd}
M.~Cirelli, A.~Strumia, and M.~Tamburini, {\em {Cosmology and Astrophysics of
  Minimal Dark Matter}},
  \href{https://dx.doi.org/10.1016/j.nuclphysb.2007.07.023}{Nucl.\  Phys.\
  {\bfseries B787} (2007) 152--175}
{\ttfamily [\href{https://arxiv.org/abs/0706.4071}{arXiv:0706.4071}]}.

\bibitem{Cirelli:2009uv}
M.~Cirelli and A.~Strumia, {\em {Minimal Dark Matter: Model and results}},
  \href{https://dx.doi.org/10.1088/1367-2630/11/10/105005}{New J.\  Phys.\
  {\bfseries 11} (2009) 105005}
{\ttfamily [\href{https://arxiv.org/abs/0903.3381}{arXiv:0903.3381}]}.

\bibitem{Kadastik:2009dj}
M.~Kadastik, K.~Kannike, and M.~Raidal, {\em {Matter parity as the origin of
  scalar Dark Matter}},
  \href{https://dx.doi.org/10.1103/PhysRevD.81.015002}{Phys.\  Rev.\  D
  {\bfseries 81} (2010) 015002} {\ttfamily
  [\href{https://arxiv.org/abs/0903.2475}{arXiv:0903.2475}]}.

\bibitem{Kadastik:2009cu}
M.~Kadastik, K.~Kannike, and M.~Raidal, {\em {Dark Matter as the signal of
  Grand Unification}},
  \href{https://dx.doi.org/10.1103/PhysRevD.80.085020}{Phys.\  Rev.\  D
  {\bfseries 80} (2009) 085020} {\ttfamily
  [\href{https://arxiv.org/abs/0907.1894}{arXiv:0907.1894}]}. [Erratum:
  Phys.Rev.D 81, 029903 (2010)].

\bibitem{Frigerio:2009wf}
M.~Frigerio and T.~Hambye, {\em {Dark matter stability and unification without
  supersymmetry}}, \href{https://dx.doi.org/10.1103/PhysRevD.81.075002}{Phys.\
  Rev.\  D {\bfseries 81} (2010) 075002} {\ttfamily
  [\href{https://arxiv.org/abs/0912.1545}{arXiv:0912.1545}]}.

\bibitem{Hambye:2010zb}
T.~Hambye, {\em {On the stability of particle dark matter}},
  \href{https://dx.doi.org/10.22323/1.110.0098}{PoS {\bfseries IDM2010} (2011)
  098} {\ttfamily [\href{https://arxiv.org/abs/1012.4587}{arXiv:1012.4587}]}.

\bibitem{Mambrini:2013iaa}
Y.~Mambrini, K.~A.~Olive, J.~Quevillon, and B.~Zaldivar, {\em {Gauge Coupling
  Unification and Nonequilibrium Thermal Dark Matter}},
  \href{https://dx.doi.org/10.1103/PhysRevLett.110.241306}{Phys.\  Rev.\
  Lett.\  {\bfseries 110} (2013) 241306} {\ttfamily
  [\href{https://arxiv.org/abs/1302.4438}{arXiv:1302.4438}]}.

\bibitem{Mambrini:2015vna}
Y.~Mambrini, N.~Nagata, K.~A.~Olive, J.~Quevillon, and J.~Zheng, {\em {Dark
  matter and gauge coupling unification in nonsupersymmetric SO(10) grand
  unified models}}, \href{https://dx.doi.org/10.1103/PhysRevD.91.095010}{Phys.\
   Rev.\  D {\bfseries 91} (2015) 095010} {\ttfamily
  [\href{https://arxiv.org/abs/1502.06929}{arXiv:1502.06929}]}.

\bibitem{Nagata:2015dma}
N.~Nagata, K.~A.~Olive, and J.~Zheng, {\em {Weakly-Interacting Massive
  Particles in Non-supersymmetric SO(10) Grand Unified Models}},
  \href{https://dx.doi.org/10.1007/JHEP10(2015)193}{JHEP {\bfseries 10} (2015)
  193}
{\ttfamily [\href{https://arxiv.org/abs/1509.00809}{arXiv:1509.00809}]}.

\bibitem{Arbelaez:2015ila}
C.~Arbelaez, R.~Longas, D.~Restrepo, and O.~Zapata, {\em {Fermion dark matter
  from SO(10) GUTs}},
  \href{https://dx.doi.org/10.1103/PhysRevD.93.013012}{Phys.\  Rev.\
  {\bfseries D93} (2016) 013012}
{\ttfamily [\href{https://arxiv.org/abs/1509.06313}{arXiv:1509.06313}]}.

\bibitem{Boucenna:2015sdg}
S.~M.~Boucenna, M.~B.~Krauss, and E.~Nardi, {\em {Dark matter from the vector
  of SO (10)}}, \href{https://dx.doi.org/10.1016/j.physletb.2016.02.008}{Phys.\
   Lett.\  {\bfseries B755} (2016) 168--176}
{\ttfamily [\href{https://arxiv.org/abs/1511.02524}{arXiv:1511.02524}]}.

\bibitem{Evans:2015cqq}
J.~L.~Evans, N.~Nagata, K.~A.~Olive, and J.~Zheng, {\em {The ATLAS Diboson
  Resonance in Non-Supersymmetric SO(10)}},
  \href{https://dx.doi.org/10.1007/JHEP02(2016)120}{JHEP {\bfseries 02} (2016)
  120}
{\ttfamily [\href{https://arxiv.org/abs/1512.02184}{arXiv:1512.02184}]}.

\bibitem{Mambrini:2016dca}
Y.~Mambrini, N.~Nagata, K.~A.~Olive, and J.~Zheng, {\em {Vacuum Stability and
  Radiative Electroweak Symmetry Breaking in an SO(10) Dark Matter Model}},
  \href{https://dx.doi.org/10.1103/PhysRevD.93.111703}{Phys.\  Rev.\
  {\bfseries D93} (2016) 111703}
{\ttfamily [\href{https://arxiv.org/abs/1602.05583}{arXiv:1602.05583}]}.

\bibitem{Parida:2016hln}
M.~K.~Parida, B.~P.~Nayak, R.~Satpathy, and R.~L.~Awasthi, {\em {Standard
  Coupling Unification in SO(10), Hybrid Seesaw Neutrino Mass and Leptogenesis,
  Dark Matter, and Proton Lifetime Predictions}},
  \href{https://dx.doi.org/10.1007/JHEP04(2017)075}{JHEP {\bfseries 04} (2017)
  075}
{\ttfamily [\href{https://arxiv.org/abs/1608.03956}{arXiv:1608.03956}]}.

\bibitem{Nagata:2016knk}
N.~Nagata, K.~A.~Olive, and J.~Zheng, {\em {Asymmetric Dark Matter Models in
  SO(10)}}, \href{https://dx.doi.org/10.1088/1475-7516/2017/02/016}{JCAP
  {\bfseries 1702} (2017) 016}
{\ttfamily [\href{https://arxiv.org/abs/1611.04693}{arXiv:1611.04693}]}.

\bibitem{Schwichtenberg:2017xhv}
J.~Schwichtenberg, {\em {Dark matter in E$_{6}$ Grand unification}},
  \href{https://dx.doi.org/10.1007/JHEP02(2018)016}{JHEP {\bfseries 02} (2018)
  016}
{\ttfamily [\href{https://arxiv.org/abs/1704.04219}{arXiv:1704.04219}]}.

\bibitem{Ma:2018uss}
E.~Ma, {\em {$SO(10) \to SU(5) \times U(1)_\chi$ as the Origin of Dark
  Matter}}, \href{https://dx.doi.org/10.1103/PhysRevD.98.091701}{Phys.\  Rev.\
  {\bfseries D98} (2018) 091701}
{\ttfamily [\href{https://arxiv.org/abs/1809.03974}{arXiv:1809.03974}]}.

\bibitem{Ferrari:2018rey}
S.~Ferrari, T.~Hambye, J.~Heeck, and M.~H.~G.~Tytgat, {\em {SO(10) paths to
  dark matter}}, \href{https://dx.doi.org/10.1103/PhysRevD.99.055032}{Phys.\
  Rev.\  {\bfseries D99} (2019) 055032}
{\ttfamily [\href{https://arxiv.org/abs/1811.07910}{arXiv:1811.07910}]}.

\bibitem{Abe:2021byq}
Y.~Abe, T.~Toma, K.~Tsumura, and N.~Yamatsu, {\em {Pseudo-Nambu-Goldstone dark
  matter model inspired by grand unification}},
  \href{https://dx.doi.org/10.1103/PhysRevD.104.035011}{Phys.\  Rev.\  D
  {\bfseries 104} (2021) 035011} {\ttfamily
  [\href{https://arxiv.org/abs/2104.13523}{arXiv:2104.13523}]}.

\bibitem{Cho:2021yue}
G.-C.~Cho, K.~Hayami, and N.~Okada, {\em {SO(10) grand unification with minimal
  dark matter and color octet scalars}},
  \href{https://dx.doi.org/10.1103/PhysRevD.105.015027}{Phys.\  Rev.\  D
  {\bfseries 105} (2022) 015027} {\ttfamily
  [\href{https://arxiv.org/abs/2110.03884}{arXiv:2110.03884}]}.

\bibitem{Abe:2020mph}
T.~Abe, M.~Fujiwara, J.~Hisano, and K.~Matsushita, {\em {A model of
  electroweakly interacting non-abelian vector dark matter}},
  \href{https://dx.doi.org/10.1007/JHEP07(2020)136}{JHEP {\bfseries 07} (2020)
  136} {\ttfamily [\href{https://arxiv.org/abs/2004.00884}{arXiv:2004.00884}]}.

\bibitem{Griest:1990kh}
K.~Griest and D.~Seckel, {\em {Three exceptions in the calculation of relic
  abundances}}, \href{https://dx.doi.org/10.1103/PhysRevD.43.3191}{Phys.\
  Rev.\  D {\bfseries 43} (1991) 3191--3203}.

\bibitem{Hisano:2003ec}
J.~Hisano, S.~Matsumoto, and M.~M.~Nojiri, {\em {Explosive dark matter
  annihilation}},
  \href{https://dx.doi.org/10.1103/PhysRevLett.92.031303}{Phys.\  Rev.\  Lett.\
   {\bfseries 92} (2004) 031303} {\ttfamily
  [\href{https://arxiv.org/abs/hep-ph/0307216}{hep-ph/0307216}]}.

\bibitem{Hisano:2004ds}
J.~Hisano, S.~Matsumoto, M.~M.~Nojiri, and O.~Saito, {\em {Non-perturbative
  effect on dark matter annihilation and gamma ray signature from galactic
  center}}, \href{https://dx.doi.org/10.1103/PhysRevD.71.063528}{Phys.\  Rev.\
  D {\bfseries 71} (2005) 063528} {\ttfamily
  [\href{https://arxiv.org/abs/hep-ph/0412403}{hep-ph/0412403}]}.

\bibitem{Hisano:2004pv}
J.~Hisano, S.~Matsumoto, M.~M.~Nojiri, and O.~Saito, {\em {Direct detection of
  the Wino and Higgsino-like neutralino dark matters at one-loop level}},
  \href{https://dx.doi.org/10.1103/PhysRevD.71.015007}{Phys.\  Rev.\
  {\bfseries D71} (2005) 015007}
{\ttfamily [\href{https://arxiv.org/abs/hep-ph/0407168}{hep-ph/0407168}]}.

\bibitem{Hisano:2010fy}
J.~Hisano, K.~Ishiwata, and N.~Nagata, {\em {A complete calculation for direct
  detection of Wino dark matter}},
  \href{https://dx.doi.org/10.1016/j.physletb.2010.05.047}{Phys.\  Lett.\
  {\bfseries B690} (2010) 311--315}
{\ttfamily [\href{https://arxiv.org/abs/1004.4090}{arXiv:1004.4090}]}.

\bibitem{Hisano:2010ct}
J.~Hisano, K.~Ishiwata, and N.~Nagata, {\em {Gluon contribution to the dark
  matter direct detection}},
  \href{https://dx.doi.org/10.1103/PhysRevD.82.115007}{Phys.\  Rev.\
  {\bfseries D82} (2010) 115007}
{\ttfamily [\href{https://arxiv.org/abs/1007.2601}{arXiv:1007.2601}]}.

\bibitem{Hisano:2011cs}
J.~Hisano, K.~Ishiwata, N.~Nagata, and T.~Takesako, {\em {Direct Detection of
  Electroweak-Interacting Dark Matter}},
  \href{https://dx.doi.org/10.1007/JHEP07(2011)005}{JHEP {\bfseries 07} (2011)
  005}
{\ttfamily [\href{https://arxiv.org/abs/1104.0228}{arXiv:1104.0228}]}.

\bibitem{Hill:2013hoa}
R.~J.~Hill and M.~P.~Solon, {\em {WIMP-nucleon scattering with heavy WIMP
  effective theory}},
  \href{https://dx.doi.org/10.1103/PhysRevLett.112.211602}{Phys.\  Rev.\
  Lett.\  {\bfseries 112} (2014) 211602}
{\ttfamily [\href{https://arxiv.org/abs/1309.4092}{arXiv:1309.4092}]}.

\bibitem{Hill:2014yxa}
R.~J.~Hill and M.~P.~Solon, {\em {Standard Model anatomy of WIMP dark matter
  direct detection II: QCD analysis and hadronic matrix elements}},
  \href{https://dx.doi.org/10.1103/PhysRevD.91.043505}{Phys.\  Rev.\
  {\bfseries D91} (2015) 043505}
{\ttfamily [\href{https://arxiv.org/abs/1409.8290}{arXiv:1409.8290}]}.

\bibitem{Hisano:2015rsa}
J.~Hisano, K.~Ishiwata, and N.~Nagata, {\em {QCD Effects on Direct Detection of
  Wino Dark Matter}}, \href{https://dx.doi.org/10.1007/JHEP06(2015)097}{JHEP
  {\bfseries 06} (2015) 097}
{\ttfamily [\href{https://arxiv.org/abs/1504.00915}{arXiv:1504.00915}]}.

\bibitem{Billard:2021uyg}
J.~Billard {\em et~al.}, {\em {Direct Detection of Dark Matter -- APPEC
  Committee Report}}, {\ttfamily
  \href{https://arxiv.org/abs/2104.07634}{arXiv:2104.07634}} (2021).

\bibitem{Kouvaris:2007ay}
C.~Kouvaris, {\em {WIMP Annihilation and Cooling of Neutron Stars}},
  \href{https://dx.doi.org/10.1103/PhysRevD.77.023006}{Phys.\  Rev.\  D
  {\bfseries 77} (2008) 023006} {\ttfamily
  [\href{https://arxiv.org/abs/0708.2362}{arXiv:0708.2362}]}.

\bibitem{Bertone:2007ae}
G.~Bertone and M.~Fairbairn, {\em {Compact Stars as Dark Matter Probes}},
  \href{https://dx.doi.org/10.1103/PhysRevD.77.043515}{Phys.\  Rev.\  D
  {\bfseries 77} (2008) 043515} {\ttfamily
  [\href{https://arxiv.org/abs/0709.1485}{arXiv:0709.1485}]}.

\bibitem{Kouvaris:2010vv}
C.~Kouvaris and P.~Tinyakov, {\em {Can Neutron stars constrain Dark Matter?}},
  \href{https://dx.doi.org/10.1103/PhysRevD.82.063531}{Phys.\  Rev.\  D
  {\bfseries 82} (2010) 063531} {\ttfamily
  [\href{https://arxiv.org/abs/1004.0586}{arXiv:1004.0586}]}.

\bibitem{deLavallaz:2010wp}
A.~de~Lavallaz and M.~Fairbairn, {\em {Neutron Stars as Dark Matter Probes}},
  \href{https://dx.doi.org/10.1103/PhysRevD.81.123521}{Phys.\  Rev.\  D
  {\bfseries 81} (2010) 123521} {\ttfamily
  [\href{https://arxiv.org/abs/1004.0629}{arXiv:1004.0629}]}.

\bibitem{Baryakhtar:2017dbj}
M.~Baryakhtar, J.~Bramante, S.~W.~Li, T.~Linden, and N.~Raj, {\em {Dark Kinetic
  Heating of Neutron Stars and An Infrared Window On WIMPs, SIMPs, and Pure
  Higgsinos}}, \href{https://dx.doi.org/10.1103/PhysRevLett.119.131801}{Phys.\
  Rev.\  Lett.\  {\bfseries 119} (2017) 131801} {\ttfamily
  [\href{https://arxiv.org/abs/1704.01577}{arXiv:1704.01577}]}.

\bibitem{Han:2021ekd}
J.~L.~Han {\em et~al.}, {\em {The FAST Galactic Plane Pulsar Snapshot survey:
  I. Project design and pulsar discoveries}},
  \href{https://dx.doi.org/10.1088/1674-4527/21/5/107}{Res.\  Astron.\
  Astrophys.\  {\bfseries 21} (2021) 107} {\ttfamily
  [\href{https://arxiv.org/abs/2105.08460}{arXiv:2105.08460}]}.

\bibitem{Gardner:2006ky}
J.~P.~Gardner {\em et~al.}, {\em {The James Webb Space Telescope}},
  \href{https://dx.doi.org/10.1007/s11214-006-8315-7}{Space Sci.\  Rev.\
  {\bfseries 123} (2006) 485} {\ttfamily
  [\href{https://arxiv.org/abs/astro-ph/0606175}{astro-ph/0606175}]}.

\bibitem{Bramante:2017xlb}
J.~Bramante, A.~Delgado, and A.~Martin, {\em {Multiscatter stellar capture of
  dark matter}}, \href{https://dx.doi.org/10.1103/PhysRevD.96.063002}{Phys.\
  Rev.\  D {\bfseries 96} (2017) 063002} {\ttfamily
  [\href{https://arxiv.org/abs/1703.04043}{arXiv:1703.04043}]}.

\bibitem{Raj:2017wrv}
N.~Raj, P.~Tanedo, and H.-B.~Yu, {\em {Neutron stars at the dark matter direct
  detection frontier}},
  \href{https://dx.doi.org/10.1103/PhysRevD.97.043006}{Phys.\  Rev.\  D
  {\bfseries 97} (2018) 043006} {\ttfamily
  [\href{https://arxiv.org/abs/1707.09442}{arXiv:1707.09442}]}.

\bibitem{Chen:2018ohx}
C.-S.~Chen and Y.-H.~Lin, {\em {Reheating neutron stars with the annihilation
  of self-interacting dark matter}},
  \href{https://dx.doi.org/10.1007/JHEP08(2018)069}{JHEP {\bfseries 08} (2018)
  069} {\ttfamily [\href{https://arxiv.org/abs/1804.03409}{arXiv:1804.03409}]}.

\bibitem{Bell:2018pkk}
N.~F.~Bell, G.~Busoni, and S.~Robles, {\em {Heating up Neutron Stars with
  Inelastic Dark Matter}},
  \href{https://dx.doi.org/10.1088/1475-7516/2018/09/018}{JCAP {\bfseries 09}
  (2018) 018} {\ttfamily
  [\href{https://arxiv.org/abs/1807.02840}{arXiv:1807.02840}]}.

\bibitem{Garani:2018kkd}
R.~Garani, Y.~Genolini, and T.~Hambye, {\em {New Analysis of Neutron Star
  Constraints on Asymmetric Dark Matter}},
  \href{https://dx.doi.org/10.1088/1475-7516/2019/05/035}{JCAP {\bfseries 05}
  (2019) 035} {\ttfamily
  [\href{https://arxiv.org/abs/1812.08773}{arXiv:1812.08773}]}.

\bibitem{Camargo:2019wou}
D.~A.~Camargo, F.~S.~Queiroz, and R.~Sturani, {\em {Detecting Dark Matter with
  Neutron Star Spectroscopy}},
  \href{https://dx.doi.org/10.1088/1475-7516/2019/09/051}{JCAP {\bfseries 09}
  (2019) 051} {\ttfamily
  [\href{https://arxiv.org/abs/1901.05474}{arXiv:1901.05474}]}.

\bibitem{Bell:2019pyc}
N.~F.~Bell, G.~Busoni, and S.~Robles, {\em {Capture of Leptophilic Dark Matter
  in Neutron Stars}},
  \href{https://dx.doi.org/10.1088/1475-7516/2019/06/054}{JCAP {\bfseries 06}
  (2019) 054} {\ttfamily
  [\href{https://arxiv.org/abs/1904.09803}{arXiv:1904.09803}]}.

\bibitem{Hamaguchi:2019oev}
K.~Hamaguchi, N.~Nagata, and K.~Yanagi, {\em {Dark Matter Heating vs.
  Rotochemical Heating in Old Neutron Stars}},
  \href{https://dx.doi.org/10.1016/j.physletb.2019.06.060}{Phys.\  Lett.\  B
  {\bfseries 795} (2019) 484--489} {\ttfamily
  [\href{https://arxiv.org/abs/1905.02991}{arXiv:1905.02991}]}.

\bibitem{Garani:2019fpa}
R.~Garani and J.~Heeck, {\em {Dark matter interactions with muons in neutron
  stars}}, \href{https://dx.doi.org/10.1103/PhysRevD.100.035039}{Phys.\  Rev.\
  D {\bfseries 100} (2019) 035039} {\ttfamily
  [\href{https://arxiv.org/abs/1906.10145}{arXiv:1906.10145}]}.

\bibitem{Acevedo:2019agu}
J.~F.~Acevedo, J.~Bramante, R.~K.~Leane, and N.~Raj, {\em {Warming Nuclear
  Pasta with Dark Matter: Kinetic and Annihilation Heating of Neutron Star
  Crusts}}, \href{https://dx.doi.org/10.1088/1475-7516/2020/03/038}{JCAP
  {\bfseries 03} (2020) 038} {\ttfamily
  [\href{https://arxiv.org/abs/1911.06334}{arXiv:1911.06334}]}.

\bibitem{Joglekar:2019vzy}
A.~Joglekar, N.~Raj, P.~Tanedo, and H.-B.~Yu, {\em {Relativistic capture of
  dark matter by electrons in neutron stars}},
  \href{https://dx.doi.org/10.1016/j.physletb.2020.135767}{Phys.\  Lett.\
  {\bfseries B} (2020) 135767} {\ttfamily
  [\href{https://arxiv.org/abs/1911.13293}{arXiv:1911.13293}]}.

\bibitem{Keung:2020teb}
W.-Y.~Keung, D.~Marfatia, and P.-Y.~Tseng, {\em {Heating neutron stars with GeV
  dark matter}}, \href{https://dx.doi.org/10.1007/JHEP07(2020)181}{JHEP
  {\bfseries 07} (2020) 181} {\ttfamily
  [\href{https://arxiv.org/abs/2001.09140}{arXiv:2001.09140}]}.

\bibitem{Yanagi:2020yvg}
K.~Yanagi, {\em {Thermal Evolution of Neutron Stars as a Probe of Physics
  beyond the Standard Model}}, {\ttfamily
  \href{https://arxiv.org/abs/2003.08199}{arXiv:2003.08199}} (2020).

\bibitem{Joglekar:2020liw}
A.~Joglekar, N.~Raj, P.~Tanedo, and H.-B.~Yu, {\em {Dark kinetic heating of
  neutron stars from contact interactions with relativistic targets}},
  \href{https://dx.doi.org/10.1103/PhysRevD.102.123002}{Phys.\  Rev.\  D
  {\bfseries 102} (2020) 123002} {\ttfamily
  [\href{https://arxiv.org/abs/2004.09539}{arXiv:2004.09539}]}.

\bibitem{Bell:2020jou}
N.~F.~Bell, G.~Busoni, S.~Robles, and M.~Virgato, {\em {Improved Treatment of
  Dark Matter Capture in Neutron Stars}},
  \href{https://dx.doi.org/10.1088/1475-7516/2020/09/028}{JCAP {\bfseries 09}
  (2020) 028} {\ttfamily
  [\href{https://arxiv.org/abs/2004.14888}{arXiv:2004.14888}]}.

\bibitem{Ilie:2020vec}
C.~Ilie, J.~Pilawa, and S.~Zhang, {\em {Comment on
  \textquotedblleft{}Multiscatter stellar capture of dark
  matter\textquotedblright{}}},
  \href{https://dx.doi.org/10.1103/PhysRevD.102.048301}{Phys.\  Rev.\  D
  {\bfseries 102} (2020) 048301} {\ttfamily
  [\href{https://arxiv.org/abs/2005.05946}{arXiv:2005.05946}]}.

\bibitem{Bell:2020lmm}
N.~F.~Bell, G.~Busoni, S.~Robles, and M.~Virgato, {\em {Improved Treatment of
  Dark Matter Capture in Neutron Stars II: Leptonic Targets}},
  \href{https://dx.doi.org/10.1088/1475-7516/2021/03/086}{JCAP {\bfseries 03}
  (2021) 086} {\ttfamily
  [\href{https://arxiv.org/abs/2010.13257}{arXiv:2010.13257}]}.

\bibitem{Bell:2020obw}
N.~F.~Bell, {\em et al.}, {\em {Nucleon Structure and Strong Interactions in
  Dark Matter Capture in Neutron Stars}},
  \href{https://dx.doi.org/10.1103/PhysRevLett.127.111803}{Phys.\  Rev.\
  Lett.\  {\bfseries 127} (2021) 111803} {\ttfamily
  [\href{https://arxiv.org/abs/2012.08918}{arXiv:2012.08918}]}.

\bibitem{Maity:2021fxw}
T.~N.~Maity and F.~S.~Queiroz, {\em {Detecting bosonic dark matter with neutron
  stars}}, \href{https://dx.doi.org/10.1103/PhysRevD.104.083019}{Phys.\  Rev.\
  D {\bfseries 104} (2021) 083019} {\ttfamily
  [\href{https://arxiv.org/abs/2104.02700}{arXiv:2104.02700}]}.

\bibitem{Anzuini:2021lnv}
F.~Anzuini, {\em et al.}, {\em {Improved treatment of dark matter capture in
  neutron stars III: nucleon and exotic targets}},
  \href{https://dx.doi.org/10.1088/1475-7516/2021/11/056}{JCAP {\bfseries 11}
  (2021) 056} {\ttfamily
  [\href{https://arxiv.org/abs/2108.02525}{arXiv:2108.02525}]}.

\bibitem{Zeng:2021moz}
Y.-P.~Zeng, X.~Xiao, and W.~Wang, {\em {Constraints on Pseudo-Nambu-Goldstone
  dark matter from direct detection experiment and neutron star reheating
  temperature}},
  \href{https://dx.doi.org/10.1016/j.physletb.2021.136822}{Phys.\  Lett.\  B
  {\bfseries 824} (2022) 136822} {\ttfamily
  [\href{https://arxiv.org/abs/2108.11381}{arXiv:2108.11381}]}.

\bibitem{Bramante:2021dyx}
J.~Bramante, B.~J.~Kavanagh, and N.~Raj, {\em {Scattering searches for dark
  matter in subhalos: neutron stars, cosmic rays, and old rocks}}, {\ttfamily
  \href{https://arxiv.org/abs/2109.04582}{arXiv:2109.04582}} (2021).

\bibitem{Tinyakov:2021lnt}
P.~Tinyakov, M.~Pshirkov, and S.~Popov, {\em {Astroparticle Physics with
  Compact Objects}}, \href{https://dx.doi.org/10.3390/universe7110401}{Universe
  {\bfseries 7} (2021) 401} {\ttfamily
  [\href{https://arxiv.org/abs/2110.12298}{arXiv:2110.12298}]}.

\bibitem{Hamaguchi:2022wpz}
K.~Hamaguchi, N.~Nagata, and M.~E.~Ramirez-Quezada, {\em {Neutron Star Heating
  in Dark Matter Models for the Muon $g-2$ Discrepancy}}, {\ttfamily
  \href{https://arxiv.org/abs/2204.02413}{arXiv:2204.02413}} (2022).

\bibitem{Hisano:2014kua}
J.~Hisano, D.~Kobayashi, N.~Mori, and E.~Senaha, {\em {Effective Interaction of
  Electroweak-Interacting Dark Matter with Higgs Boson and Its Phenomenology}},
  \href{https://dx.doi.org/10.1016/j.physletb.2015.01.012}{Phys.\  Lett.\  B
  {\bfseries 742} (2015) 80--85} {\ttfamily
  [\href{https://arxiv.org/abs/1410.3569}{arXiv:1410.3569}]}.

\bibitem{Nagata:2014wma}
N.~Nagata and S.~Shirai, {\em {Higgsino Dark Matter in High-Scale
  Supersymmetry}}, \href{https://dx.doi.org/10.1007/JHEP01(2015)029}{JHEP
  {\bfseries 01} (2015) 029} {\ttfamily
  [\href{https://arxiv.org/abs/1410.4549}{arXiv:1410.4549}]}.

\bibitem{Nagata:2014aoa}
N.~Nagata and S.~Shirai, {\em {Electroweakly-Interacting Dirac Dark Matter}},
  \href{https://dx.doi.org/10.1103/PhysRevD.91.055035}{Phys.\  Rev.\  D
  {\bfseries 91} (2015) 055035} {\ttfamily
  [\href{https://arxiv.org/abs/1411.0752}{arXiv:1411.0752}]}.

\bibitem{Dedes:2016odh}
A.~Dedes, D.~Karamitros, and V.~C.~Spanos, {\em {Effective Theory for
  Electroweak Doublet Dark Matter}},
  \href{https://dx.doi.org/10.1103/PhysRevD.94.095008}{Phys.\  Rev.\  D
  {\bfseries 94} (2016) 095008} {\ttfamily
  [\href{https://arxiv.org/abs/1607.05040}{arXiv:1607.05040}]}.

\bibitem{Fukuda:2017jmk}
H.~Fukuda, N.~Nagata, H.~Otono, and S.~Shirai, {\em {Higgsino Dark Matter or
  Not: Role of Disappearing Track Searches at the LHC and Future Colliders}},
  \href{https://dx.doi.org/10.1016/j.physletb.2018.03.088}{Phys.\  Lett.\  B
  {\bfseries 781} (2018) 306--311} {\ttfamily
  [\href{https://arxiv.org/abs/1703.09675}{arXiv:1703.09675}]}.

\bibitem{Kuramoto:2019yvj}
W.~Kuramoto, T.~Kuwahara, and R.~Nagai, {\em {Renormalization Effects on
  Electric Dipole Moments in Electroweakly Interacting Massive Particle
  Models}}, \href{https://dx.doi.org/10.1103/PhysRevD.99.095024}{Phys.\  Rev.\
  D {\bfseries 99} (2019) 095024} {\ttfamily
  [\href{https://arxiv.org/abs/1902.05360}{arXiv:1902.05360}]}.

\bibitem{Geytenbeek:2020rxa}
B.~Geytenbeek and B.~Gripaios, {\em {Effective field theory analysis of
  composite higgsino-like and wino-like thermal relic dark matter}},
  \href{https://dx.doi.org/10.1088/1475-7516/2021/05/060}{JCAP {\bfseries 05}
  (2021) 060} {\ttfamily
  [\href{https://arxiv.org/abs/2011.06025}{arXiv:2011.06025}]}.

\bibitem{Belyaev:2022qnf}
A.~Belyaev, G.~Cacciapaglia, D.~Locke, and A.~Pukhov, {\em {Minimal Consistent
  Dark Matter models for systematic experimental characterisation: Fermion Dark
  Matter}}, {\ttfamily
  \href{https://arxiv.org/abs/2203.03660}{arXiv:2203.03660}} (2022).

\bibitem{Goldman:1989nd}
I.~Goldman and S.~Nussinov, {\em {Weakly Interacting Massive Particles and
  Neutron Stars}}, \href{https://dx.doi.org/10.1103/PhysRevD.40.3221}{Phys.\
  Rev.\  D {\bfseries 40} (1989) 3221--3230}.

\bibitem{Akmal:1998cf}
A.~Akmal, V.~R.~Pandharipande, and D.~G.~Ravenhall, {\em {The Equation of state
  of nucleon matter and neutron star structure}},
  \href{https://dx.doi.org/10.1103/PhysRevC.58.1804}{Phys.\  Rev.\  C
  {\bfseries 58} (1998) 1804--1828} {\ttfamily
  [\href{https://arxiv.org/abs/nucl-th/9804027}{nucl-th/9804027}]}.

\bibitem{Li:2018lpy}
B.-A.~Li, B.-J.~Cai, L.-W.~Chen, and J.~Xu, {\em {Nucleon Effective Masses in
  Neutron-Rich Matter}},
  \href{https://dx.doi.org/10.1016/j.ppnp.2018.01.001}{Prog.\  Part.\  Nucl.\
  Phys.\  {\bfseries 99} (2018) 29--119} {\ttfamily
  [\href{https://arxiv.org/abs/1801.01213}{arXiv:1801.01213}]}.

\bibitem{Bertoni:2013bsa}
B.~Bertoni, A.~E.~Nelson, and S.~Reddy, {\em {Dark Matter Thermalization in
  Neutron Stars}}, \href{https://dx.doi.org/10.1103/PhysRevD.88.123505}{Phys.\
  Rev.\  D {\bfseries 88} (2013) 123505} {\ttfamily
  [\href{https://arxiv.org/abs/1309.1721}{arXiv:1309.1721}]}.

\bibitem{Garani:2020wge}
R.~Garani, A.~Gupta, and N.~Raj, {\em {Observing the thermalization of dark
  matter in neutron stars}},
  \href{https://dx.doi.org/10.1103/PhysRevD.103.043019}{Phys.\  Rev.\  D
  {\bfseries 103} (2021) 043019} {\ttfamily
  [\href{https://arxiv.org/abs/2009.10728}{arXiv:2009.10728}]}.

\bibitem{Yakovlev:1999sk}
D.~G.~Yakovlev, K.~P.~Levenfish, and Y.~A.~Shibanov, {\em {Cooling neutron
  stars and superfluidity in their interiors}},
  \href{https://dx.doi.org/10.1070/PU1999v042n08ABEH000556}{Phys.\  Usp.\
  {\bfseries 42} (1999) 737--778} {\ttfamily
  [\href{https://arxiv.org/abs/astro-ph/9906456}{astro-ph/9906456}]}.

\bibitem{Yakovlev:2000jp}
D.~G.~Yakovlev, A.~D.~Kaminker, O.~Y.~Gnedin, and P.~Haensel, {\em {Neutrino
  emission from neutron stars}},
  \href{https://dx.doi.org/10.1016/S0370-1573(00)00131-9}{Phys.\  Rept.\
  {\bfseries 354} (2001) 1} {\ttfamily
  [\href{https://arxiv.org/abs/astro-ph/0012122}{astro-ph/0012122}]}.

\bibitem{Yakovlev:2004iq}
D.~G.~Yakovlev and C.~J.~Pethick, {\em {Neutron star cooling}},
  \href{https://dx.doi.org/10.1146/annurev.astro.42.053102.134013}{Ann.\  Rev.\
   Astron.\  Astrophys.\  {\bfseries 42} (2004) 169--210} {\ttfamily
  [\href{https://arxiv.org/abs/astro-ph/0402143}{astro-ph/0402143}]}.

\bibitem{Page:2004fy}
D.~Page, J.~M.~Lattimer, M.~Prakash, and A.~W.~Steiner, {\em {Minimal cooling
  of neutron stars: A New paradigm}},
  \href{https://dx.doi.org/10.1086/424844}{Astrophys.\  J.\  Suppl.\
  {\bfseries 155} (2004) 623--650} {\ttfamily
  [\href{https://arxiv.org/abs/astro-ph/0403657}{astro-ph/0403657}]}.

\bibitem{Page:2009fu}
D.~Page, J.~M.~Lattimer, M.~Prakash, and A.~W.~Steiner, {\em {Neutrino Emission
  from Cooper Pairs and Minimal Cooling of Neutron Stars}},
  \href{https://dx.doi.org/10.1088/0004-637X/707/2/1131}{Astrophys.\  J.\
  {\bfseries 707} (2009) 1131--1140} {\ttfamily
  [\href{https://arxiv.org/abs/0906.1621}{arXiv:0906.1621}]}.

\bibitem{Potekhin:2015qsa}
A.~Y.~Potekhin, J.~A.~Pons, and D.~Page, {\em {Neutron stars - cooling and
  transport}}, \href{https://dx.doi.org/10.1007/s11214-015-0180-9}{Space Sci.\
  Rev.\  {\bfseries 191} (2015) 239--291} {\ttfamily
  [\href{https://arxiv.org/abs/1507.06186}{arXiv:1507.06186}]}.

\bibitem{tempdata}
{\em {Cooling neutron stars}},
  \url{http://www.ioffe.ru/astro/NSG/thermal/cooldat.html}.

\bibitem{Potekhin:2020ttj}
A.~Y.~Potekhin, D.~A.~Zyuzin, D.~G.~Yakovlev, M.~V.~Beznogov, and
  Y.~A.~Shibanov, {\em {Thermal luminosities of cooling neutron stars}},
  \href{https://dx.doi.org/10.1093/mnras/staa1871}{Mon.\  Not.\  Roy.\
  Astron.\  Soc.\  {\bfseries 496} (2020) 5052--5071} {\ttfamily
  [\href{https://arxiv.org/abs/2006.15004}{arXiv:2006.15004}]}.

\bibitem{Kargaltsev:2003eb}
O.~Kargaltsev, G.~G.~Pavlov, and R.~W.~Romani, {\em {Ultraviolet emission from
  the millisecond pulsar j0437-4715}},
  \href{https://dx.doi.org/10.1086/380993}{Astrophys.\  J.\  {\bfseries 602}
  (2004) 327--335} {\ttfamily
  [\href{https://arxiv.org/abs/astro-ph/0310854}{astro-ph/0310854}]}.

\bibitem{Mignani:2008jr}
R.~P.~Mignani, G.~G.~Pavlov, and O.~Kargaltsev, {\em {A possible optical
  counterpart to the old nearby pulsar J0108-1431}},
  \href{https://dx.doi.org/10.1051/0004-6361:200810212}{Astron.\  Astrophys.\
  {\bfseries 488} (2008) 1027} {\ttfamily
  [\href{https://arxiv.org/abs/0805.2586}{arXiv:0805.2586}]}.

\bibitem{Durant:2011je}
M.~Durant, {\em et al.}, {\em {The spectrum of the recycled PSR J0437-4715 and
  its white dwarf companion}},
  \href{https://dx.doi.org/10.1088/0004-637X/746/1/6}{Astrophys.\  J.\
  {\bfseries 746} (2012) 6} {\ttfamily
  [\href{https://arxiv.org/abs/1111.2346}{arXiv:1111.2346}]}.

\bibitem{Rangelov:2016syg}
B.~Rangelov, {\em et al.}, {\em {Hubble Space Telescope Detection of the
  Millisecond Pulsar J2124\ensuremath{-}3358 and its Far-ultraviolet Bow Shock
  Nebula}}, \href{https://dx.doi.org/10.3847/1538-4357/835/2/264}{Astrophys.\
  J.\  {\bfseries 835} (2017) 264} {\ttfamily
  [\href{https://arxiv.org/abs/1701.00002}{arXiv:1701.00002}]}.

\bibitem{Pavlov:2017eeu}
G.~G.~Pavlov, {\em et al.}, {\em {Old but still warm: Far-UV detection of PSR
  B0950+08}}, \href{https://dx.doi.org/10.3847/1538-4357/aa947c}{Astrophys.\
  J.\  {\bfseries 850} (2017) 79} {\ttfamily
  [\href{https://arxiv.org/abs/1710.06448}{arXiv:1710.06448}]}.

\bibitem{Abramkin:2021fzy}
V.~Abramkin, G.~G.~Pavlov, Y.~Shibanov, and O.~Kargaltsev, {\em {Thermal and
  Nonthermal Emission in the Optical-UV Spectrum of PSR B0950+08*}},
  \href{https://dx.doi.org/10.3847/1538-4357/ac3a6f}{Astrophys.\  J.\
  {\bfseries 924} (2022) 128} {\ttfamily
  [\href{https://arxiv.org/abs/2111.08801}{arXiv:2111.08801}]}.

\bibitem{Yanagi:2019vrr}
K.~Yanagi, N.~Nagata, and K.~Hamaguchi, {\em {Cooling Theory Faced with Old
  Warm Neutron Stars: Role of Non-Equilibrium Processes with Proton and Neutron
  Gaps}}, \href{https://dx.doi.org/10.1093/mnras/staa076}{Mon.\  Not.\  Roy.\
  Astron.\  Soc.\  {\bfseries 492} (2020) 5508--5523} {\ttfamily
  [\href{https://arxiv.org/abs/1904.04667}{arXiv:1904.04667}]}.

\bibitem{Gonzalez:2010ta}
D.~Gonzalez and A.~Reisenegger, {\em {Internal Heating of Old Neutron Stars:
  Contrasting Different Mechanisms}},
  \href{https://dx.doi.org/10.1051/0004-6361/201015084}{Astron.\  Astrophys.\
  {\bfseries 522} (2010) A16} {\ttfamily
  [\href{https://arxiv.org/abs/1005.5699}{arXiv:1005.5699}]}.

\bibitem{Reisenegger:1994be}
A.~Reisenegger, {\em {Deviations from chemical equilibrium due to spindown as
  an internal heat source in neutron stars}},
  \href{https://dx.doi.org/10.1086/175480}{Astrophys.\  J.\  {\bfseries 442}
  (1995) 749} {\ttfamily
  [\href{https://arxiv.org/abs/astro-ph/9410035}{astro-ph/9410035}]}.

\bibitem{1992A&A...262..131H}
P.~{Haensel}, {\em {Non-equilibrium neutrino emissivities and opacities of
  neutron star matter}}, \aap {\bfseries 262} (1992) 131--137.

\bibitem{1993A&A...271..187G}
E.~{Gourgoulhon} and P.~{Haensel}, {\em {Upper bounds on the neutrino burst
  from collapse of a neutron star into a black hole}}, \aap {\bfseries 271}
  (1993) 187.

\bibitem{Fernandez:2005cg}
R.~Fernandez and A.~Reisenegger, {\em {Rotochemical heating in millisecond
  pulsars. Formalism and non-superfluid case}},
  \href{https://dx.doi.org/10.1086/429551}{Astrophys.\  J.\  {\bfseries 625}
  (2005) 291--306} {\ttfamily
  [\href{https://arxiv.org/abs/astro-ph/0502116}{astro-ph/0502116}]}.

\bibitem{Villain:2005ns}
L.~Villain and P.~Haensel, {\em {Non-equilibrium beta processes in superfluid
  neutron star cores}},
  \href{https://dx.doi.org/10.1051/0004-6361:20053313}{Astron.\  Astrophys.\
  {\bfseries 444} (2005) 539} {\ttfamily
  [\href{https://arxiv.org/abs/astro-ph/0504572}{astro-ph/0504572}]}.

\bibitem{Petrovich:2009yh}
C.~Petrovich and A.~Reisenegger, {\em {Rotochemical heating in millisecond
  pulsars: modified Urca reactions with uniform Cooper pairing gaps}},
  \href{https://dx.doi.org/10.1051/0004-6361/200913861}{Astron.\  Astrophys.\
  {\bfseries 521} (2010) A77} {\ttfamily
  [\href{https://arxiv.org/abs/0912.2564}{arXiv:0912.2564}]}.

\bibitem{Pi:2009eq}
C.-M.~Pi, X.-P.~Zheng, and S.-H.~Yang, {\em {Neutrino Emissivity of
  Non-equilibrium beta processes With Nucleon Superfluidity}},
  \href{https://dx.doi.org/10.1103/PhysRevC.81.045802}{Phys.\  Rev.\  C
  {\bfseries 81} (2010) 045802} {\ttfamily
  [\href{https://arxiv.org/abs/0912.2884}{arXiv:0912.2884}]}.

\bibitem{Gonzalez-Jimenez:2014iia}
N.~Gonz\'alez-Jim\'enez, C.~Petrovich, and A.~Reisenegger, {\em {Rotochemical
  heating of millisecond and classical pulsars with anisotropic and
  density-dependent superfluid gap models}},
  \href{https://dx.doi.org/10.1093/mnras/stu2558}{Mon.\  Not.\  Roy.\  Astron.\
   Soc.\  {\bfseries 447} (2015) 2073} {\ttfamily
  [\href{https://arxiv.org/abs/1411.6500}{arXiv:1411.6500}]}.

\bibitem{1984ApJ...276..325A}
M.~A.~{Alpar}, D.~{Pines}, P.~W.~{Anderson}, and J.~{Shaham}, {\em {Vortex
  creep and the internal temperature of neutron stars. I - General theory}},
  \href{https://dx.doi.org/10.1086/161616}{\apj {\bfseries 276} (1984)
  325--334}.

\bibitem{1989ApJ...346..808S}
N.~{Shibazaki} and F.~K.~{Lamb}, {\em {Neutron star evolution with internal
  heating}}, \href{https://dx.doi.org/10.1086/168062}{Astrophys.\  J.\
  {\bfseries 346} (1989) 808--822}.

\bibitem{1991ApJ...381L..47V}
K.~A.~{van Riper}, R.~I.~{Epstein}, and G.~S.~{Miller}, {\em {Soft X-ray pulses
  from neutron star glitches}}, \href{https://dx.doi.org/10.1086/186193}{\apj
  {\bfseries 381} (1991) L47--L50}.

\bibitem{1993ApJ...408..186U}
H.~{Umeda}, N.~{Shibazaki}, K.~{Nomoto}, and S.~{Tsuruta}, {\em {Thermal
  evolution of neutron stars with internal frictional heating}},
  \href{https://dx.doi.org/10.1086/172579}{\apj {\bfseries 408} (1993)
  186--193}.

\bibitem{VanRiper:1994vp}
K.~Van~Riper, B.~Link, and R.~Epstein, {\em {Frictional heating and neutron
  star thermal evolution}},
  \href{https://dx.doi.org/10.1086/175961}{Astrophys.\  J.\  {\bfseries 448}
  (1995) 294}
{\ttfamily [\href{https://arxiv.org/abs/astro-ph/9404060}{astro-ph/9404060}]}.

\bibitem{Larson:1998it}
M.~B.~Larson and B.~Link, {\em {Superfluid friction and late-time thermal
  evolution of neutron stars}},
  \href{https://dx.doi.org/10.1086/307532}{Astrophys.\  J.\  {\bfseries 521}
  (1999) 271}
{\ttfamily [\href{https://arxiv.org/abs/astro-ph/9810441}{astro-ph/9810441}]}.

\bibitem{Gusakov:2015kaa}
M.~E.~Gusakov, E.~M.~Kantor, and A.~Reisenegger, {\em {Rotation-induced deep
  crustal heating of millisecond pulsars}},
  \href{https://dx.doi.org/10.1093/mnrasl/slv095}{Mon.\  Not.\  Roy.\  Astron.\
   Soc.\  {\bfseries 453} (2015) L36--L40} {\ttfamily
  [\href{https://arxiv.org/abs/1507.04586}{arXiv:1507.04586}]}.

\bibitem{Hambye:2009pw}
T.~Hambye, F.~S.~Ling, L.~Lopez~Honorez, and J.~Rocher, {\em {Scalar Multiplet
  Dark Matter}}, \href{https://dx.doi.org/10.1007/JHEP05(2010)066}{JHEP
  {\bfseries 07} (2009) 090} {\ttfamily
  [\href{https://arxiv.org/abs/0903.4010}{arXiv:0903.4010}]}. [Erratum: JHEP
  05, 066 (2010)].

\bibitem{DiLuzio:2015oha}
L.~Di~Luzio, R.~Gr^^c3^^b6ber, J.~F.~Kamenik, and M.~Nardecchia, {\em
  {Accidental matter at the LHC}},
  \href{https://dx.doi.org/10.1007/JHEP07(2015)074}{JHEP {\bfseries 07} (2015)
  074} {\ttfamily [\href{https://arxiv.org/abs/1504.00359}{arXiv:1504.00359}]}.

\bibitem{Logan:2016ivc}
H.~E.~Logan and T.~Pilkington, {\em {Large scalar multiplet dark matter in the
  high-mass region}},
  \href{https://dx.doi.org/10.1103/PhysRevD.96.015030}{Phys.\  Rev.\  D
  {\bfseries 96} (2017) 015030} {\ttfamily
  [\href{https://arxiv.org/abs/1610.08835}{arXiv:1610.08835}]}.

\bibitem{Cai:2017fmr}
C.~Cai, Z.~Kang, Z.~Luo, Z.-H.~Yu, and H.-H.~Zhang, {\em {Scalar quintuplet
  minimal dark matter with Yukawa interactions: perturbative up to the Planck
  scale}}, \href{https://dx.doi.org/10.1088/1674-1137/43/2/023102}{Chin.\
  Phys.\  C {\bfseries 43} (2019) 023102} {\ttfamily
  [\href{https://arxiv.org/abs/1711.07396}{arXiv:1711.07396}]}.

\bibitem{Belyaev:2018xpf}
A.~Belyaev, G.~Cacciapaglia, J.~Mckay, D.~Marin, and A.~R.~Zerwekh, {\em
  {Minimal Spin-one Isotriplet Dark Matter}},
  \href{https://dx.doi.org/10.1103/PhysRevD.99.115003}{Phys.\  Rev.\  D
  {\bfseries 99} (2019) 115003} {\ttfamily
  [\href{https://arxiv.org/abs/1808.10464}{arXiv:1808.10464}]}.

\bibitem{Chao:2018xwz}
W.~Chao, G.-J.~Ding, X.-G.~He, and M.~Ramsey-Musolf, {\em {Scalar Electroweak
  Multiplet Dark Matter}},
  \href{https://dx.doi.org/10.1007/JHEP08(2019)058}{JHEP {\bfseries 08} (2019)
  058} {\ttfamily [\href{https://arxiv.org/abs/1812.07829}{arXiv:1812.07829}]}.

\bibitem{Chiang:2020rcv}
C.-W.~Chiang, G.~Cottin, Y.~Du, K.~Fuyuto, and M.~J.~Ramsey-Musolf, {\em
  {Collider Probes of Real Triplet Scalar Dark Matter}},
  \href{https://dx.doi.org/10.1007/JHEP01(2021)198}{JHEP {\bfseries 01} (2021)
  198} {\ttfamily [\href{https://arxiv.org/abs/2003.07867}{arXiv:2003.07867}]}.

\bibitem{Belyaev:2020wok}
A.~Belyaev, S.~Prestel, F.~Rojas-Abbate, and J.~Zurita, {\em {Probing dark
  matter with disappearing tracks at the LHC}},
  \href{https://dx.doi.org/10.1103/PhysRevD.103.095006}{Phys.\  Rev.\  D
  {\bfseries 103} (2021) 095006} {\ttfamily
  [\href{https://arxiv.org/abs/2008.08581}{arXiv:2008.08581}]}.

\bibitem{Mizuta:1992ja}
S.~Mizuta, D.~Ng, and M.~Yamaguchi, {\em {Phenomenological aspects of
  supersymmetric standard models without grand unification}},
  \href{https://dx.doi.org/10.1016/0370-2693(93)90754-6}{Phys.\  Lett.\  B
  {\bfseries 300} (1993) 96--103} {\ttfamily
  [\href{https://arxiv.org/abs/hep-ph/9210241}{hep-ph/9210241}]}.

\bibitem{Pierce:1994ew}
D.~Pierce and A.~Papadopoulos, {\em {The Complete radiative corrections to the
  gaugino and Higgsino masses in the minimal supersymmetric model}},
  \href{https://dx.doi.org/10.1016/0550-3213(94)00303-3}{Nucl.\  Phys.\  B
  {\bfseries 430} (1994) 278--294} {\ttfamily
  [\href{https://arxiv.org/abs/hep-ph/9403240}{hep-ph/9403240}]}.

\bibitem{Cheng:1998hc}
H.-C.~Cheng, B.~A.~Dobrescu, and K.~T.~Matchev, {\em {Generic and chiral
  extensions of the supersymmetric standard model}},
  \href{https://dx.doi.org/10.1016/S0550-3213(99)00012-7}{Nucl.\  Phys.\  B
  {\bfseries 543} (1999) 47--72} {\ttfamily
  [\href{https://arxiv.org/abs/hep-ph/9811316}{hep-ph/9811316}]}.

\bibitem{Feng:1999fu}
J.~L.~Feng, T.~Moroi, L.~Randall, M.~Strassler, and S.-f.~Su, {\em {Discovering
  supersymmetry at the Tevatron in wino LSP scenarios}},
  \href{https://dx.doi.org/10.1103/PhysRevLett.83.1731}{Phys.\  Rev.\  Lett.\
  {\bfseries 83} (1999) 1731--1734} {\ttfamily
  [\href{https://arxiv.org/abs/hep-ph/9904250}{hep-ph/9904250}]}.

\bibitem{Gherghetta:1999sw}
T.~Gherghetta, G.~F.~Giudice, and J.~D.~Wells, {\em {Phenomenological
  consequences of supersymmetry with anomaly induced masses}},
  \href{https://dx.doi.org/10.1016/S0550-3213(99)00429-0}{Nucl.\  Phys.\  B
  {\bfseries 559} (1999) 27--47} {\ttfamily
  [\href{https://arxiv.org/abs/hep-ph/9904378}{hep-ph/9904378}]}.

\bibitem{Yamada:2009ve}
Y.~Yamada, {\em {Electroweak two-loop contribution to the mass splitting within
  a new heavy SU(2)(L) fermion multiplet}},
  \href{https://dx.doi.org/10.1016/j.physletb.2009.11.044}{Phys.\  Lett.\  B
  {\bfseries 682} (2010) 435--440} {\ttfamily
  [\href{https://arxiv.org/abs/0906.5207}{arXiv:0906.5207}]}.

\bibitem{Ibe:2012sx}
M.~Ibe, S.~Matsumoto, and R.~Sato, {\em {Mass Splitting between Charged and
  Neutral Winos at Two-Loop Level}},
  \href{https://dx.doi.org/10.1016/j.physletb.2013.03.015}{Phys.\  Lett.\  B
  {\bfseries 721} (2013) 252--260} {\ttfamily
  [\href{https://arxiv.org/abs/1212.5989}{arXiv:1212.5989}]}.

\bibitem{McKay:2017xlc}
J.~McKay and P.~Scott, {\em {Two-loop mass splittings in electroweak
  multiplets: winos and minimal dark matter}},
  \href{https://dx.doi.org/10.1103/PhysRevD.97.055049}{Phys.\  Rev.\  D
  {\bfseries 97} (2018) 055049} {\ttfamily
  [\href{https://arxiv.org/abs/1712.00968}{arXiv:1712.00968}]}.

\bibitem{Hisano:2006nn}
J.~Hisano, S.~Matsumoto, M.~Nagai, O.~Saito, and M.~Senami, {\em
  {Non-perturbative effect on thermal relic abundance of dark matter}},
  \href{https://dx.doi.org/10.1016/j.physletb.2007.01.012}{Phys.\  Lett.\  B
  {\bfseries 646} (2007) 34--38} {\ttfamily
  [\href{https://arxiv.org/abs/hep-ph/0610249}{hep-ph/0610249}]}.

\bibitem{Mitridate:2017izz}
A.~Mitridate, M.~Redi, J.~Smirnov, and A.~Strumia, {\em {Cosmological
  Implications of Dark Matter Bound States}},
  \href{https://dx.doi.org/10.1088/1475-7516/2017/05/006}{JCAP {\bfseries 05}
  (2017) 006} {\ttfamily
  [\href{https://arxiv.org/abs/1702.01141}{arXiv:1702.01141}]}.

\bibitem{Barr:1990vd}
S.~M.~Barr and A.~Zee, {\em {Electric Dipole Moment of the Electron and of the
  Neutron}}, \href{https://dx.doi.org/10.1103/PhysRevLett.65.21}{Phys.\  Rev.\
  Lett.\  {\bfseries 65} (1990) 21--24}. [Erratum: Phys.Rev.Lett. 65, 2920
  (1990)].

\bibitem{Farina:2013mla}
M.~Farina, D.~Pappadopulo, and A.~Strumia, {\em {A modified naturalness
  principle and its experimental tests}},
  \href{https://dx.doi.org/10.1007/JHEP08(2013)022}{JHEP {\bfseries 08} (2013)
  022} {\ttfamily [\href{https://arxiv.org/abs/1303.7244}{arXiv:1303.7244}]}.

\bibitem{Bagnasco:1993st}
J.~Bagnasco, M.~Dine, and S.~D.~Thomas, {\em {Detecting technibaryon dark
  matter}}, \href{https://dx.doi.org/10.1016/0370-2693(94)90830-3}{Phys.\
  Lett.\  B {\bfseries 320} (1994) 99--104} {\ttfamily
  [\href{https://arxiv.org/abs/hep-ph/9310290}{hep-ph/9310290}]}.

\bibitem{Sigurdson:2004zp}
K.~Sigurdson, M.~Doran, A.~Kurylov, R.~R.~Caldwell, and M.~Kamionkowski, {\em
  {Dark-matter electric and magnetic dipole moments}},
  \href{https://dx.doi.org/10.1103/PhysRevD.70.083501}{Phys.\  Rev.\  D
  {\bfseries 70} (2004) 083501} {\ttfamily
  [\href{https://arxiv.org/abs/astro-ph/0406355}{astro-ph/0406355}]}. [Erratum:
  Phys.Rev.D 73, 089903 (2006)].

\bibitem{Masso:2009mu}
E.~Masso, S.~Mohanty, and S.~Rao, {\em {Dipolar Dark Matter}},
  \href{https://dx.doi.org/10.1103/PhysRevD.80.036009}{Phys.\  Rev.\  D
  {\bfseries 80} (2009) 036009} {\ttfamily
  [\href{https://arxiv.org/abs/0906.1979}{arXiv:0906.1979}]}.

\bibitem{Cho:2010br}
W.~S.~Cho, J.-H.~Huh, I.-W.~Kim, J.~E.~Kim, and B.~Kyae, {\em {Constraining
  WIMP magnetic moment from CDMS II experiment}},
  \href{https://dx.doi.org/10.1016/j.physletb.2010.09.048}{Phys.\  Lett.\  B
  {\bfseries 687} (2010) 6--10} {\ttfamily
  [\href{https://arxiv.org/abs/1001.0579}{arXiv:1001.0579}]}. [Erratum:
  Phys.Lett.B 694, 496--497 (2011)].

\bibitem{Banks:2010eh}
T.~Banks, J.-F.~Fortin, and S.~Thomas, {\em {Direct Detection of Dark Matter
  Electromagnetic Dipole Moments}}, {\ttfamily
  \href{https://arxiv.org/abs/1007.5515}{arXiv:1007.5515}} (2010).

\bibitem{Ibarra:2015fqa}
A.~Ibarra and S.~Wild, {\em {Dirac dark matter with a charged mediator: a
  comprehensive one-loop analysis of the direct detection phenomenology}},
  \href{https://dx.doi.org/10.1088/1475-7516/2015/05/047}{JCAP {\bfseries 05}
  (2015) 047} {\ttfamily
  [\href{https://arxiv.org/abs/1503.03382}{arXiv:1503.03382}]}.

\bibitem{Herrero-Garcia:2018koq}
J.~Herrero-Garcia, E.~Molinaro, and M.~A.~Schmidt, {\em {Dark matter direct
  detection of a fermionic singlet at one loop}},
  \href{https://dx.doi.org/10.1140/epjc/s10052-018-5935-5}{Eur.\  Phys.\  J.\
  C {\bfseries 78} (2018) 471} {\ttfamily
  [\href{https://arxiv.org/abs/1803.05660}{arXiv:1803.05660}]}.

\bibitem{Hisano:2018bpz}
J.~Hisano, R.~Nagai, and N.~Nagata, {\em {Singlet Dirac Fermion Dark Matter
  with Mediators at Loop}},
  \href{https://dx.doi.org/10.1007/JHEP12(2018)059}{JHEP {\bfseries 12} (2018)
  059} {\ttfamily [\href{https://arxiv.org/abs/1808.06301}{arXiv:1808.06301}]}.

\bibitem{Feldstein:2013uha}
B.~Feldstein, M.~Ibe, and T.~T.~Yanagida, {\em {Hypercharged Dark Matter and
  Direct Detection as a Probe of Reheating}},
  \href{https://dx.doi.org/10.1103/PhysRevLett.112.101301}{Phys.\ Rev.\ Lett.\
  {\bfseries 112} (2013) 101301}
{\ttfamily [\href{https://arxiv.org/abs/1310.7495}{arXiv:1310.7495}]}.

\bibitem{PandaX-4T:2021bab}
{\bfseries PandaX-4T} Collaboration, {\em {Dark Matter Search Results from the
  PandaX-4T Commissioning Run}},
  \href{https://dx.doi.org/10.1103/PhysRevLett.127.261802}{Phys.\  Rev.\
  Lett.\  {\bfseries 127} (2021) 261802} {\ttfamily
  [\href{https://arxiv.org/abs/2107.13438}{arXiv:2107.13438}]}.

\bibitem{XENON:2019rxp}
{\bfseries XENON} Collaboration, {\em {Constraining the spin-dependent
  WIMP-nucleon cross sections with XENON1T}},
  \href{https://dx.doi.org/10.1103/PhysRevLett.122.141301}{Phys.\  Rev.\
  Lett.\  {\bfseries 122} (2019) 141301} {\ttfamily
  [\href{https://arxiv.org/abs/1902.03234}{arXiv:1902.03234}]}.

\bibitem{PICO:2019vsc}
{\bfseries PICO} Collaboration, {\em {Dark Matter Search Results from the
  Complete Exposure of the PICO-60 C$_3$F$_8$ Bubble Chamber}},
  \href{https://dx.doi.org/10.1103/PhysRevD.100.022001}{Phys.\  Rev.\  D
  {\bfseries 100} (2019) 022001} {\ttfamily
  [\href{https://arxiv.org/abs/1902.04031}{arXiv:1902.04031}]}.

\bibitem{Shifman:1978zn}
M.~A.~Shifman, A.~I.~Vainshtein, and V.~I.~Zakharov, {\em {Remarks on Higgs
  Boson Interactions with Nucleons}},
  \href{https://dx.doi.org/10.1016/0370-2693(78)90481-1}{Phys.\  Lett.\  B
  {\bfseries 78} (1978) 443--446}.

\bibitem{Ellis:2018dmb}
J.~Ellis, N.~Nagata, and K.~A.~Olive, {\em {Uncertainties in WIMP Dark Matter
  Scattering Revisited}},
  \href{https://dx.doi.org/10.1140/epjc/s10052-018-6047-y}{Eur.\  Phys.\  J.\
  C {\bfseries 78} (2018) 569} {\ttfamily
  [\href{https://arxiv.org/abs/1805.09795}{arXiv:1805.09795}]}.

\bibitem{XENON:2020kmp}
{\bfseries XENON} Collaboration, {\em {Projected WIMP sensitivity of the
  XENONnT dark matter experiment}},
  \href{https://dx.doi.org/10.1088/1475-7516/2020/11/031}{JCAP {\bfseries 11}
  (2020) 031} {\ttfamily
  [\href{https://arxiv.org/abs/2007.08796}{arXiv:2007.08796}]}.

\bibitem{Alexandrou:2019brg}
C.~Alexandrou, {\em et al.}, {\em {Nucleon axial, tensor, and scalar charges
  and $\sigma$-terms in lattice QCD}},
  \href{https://dx.doi.org/10.1103/PhysRevD.102.054517}{Phys.\  Rev.\  D
  {\bfseries 102} (2020) 054517} {\ttfamily
  [\href{https://arxiv.org/abs/1909.00485}{arXiv:1909.00485}]}.

\bibitem{IceCube:2016dgk}
{\bfseries IceCube} Collaboration, {\em {Search for annihilating dark matter in
  the Sun with 3 years of IceCube data}},
  \href{https://dx.doi.org/10.1140/epjc/s10052-017-4689-9}{Eur.\  Phys.\  J.\
  C {\bfseries 77} (2017) 146} {\ttfamily
  [\href{https://arxiv.org/abs/1612.05949}{arXiv:1612.05949}]}. [Erratum:
  Eur.Phys.J.C 79, 214 (2019)].

\end{thebibliography}\endgroup


\end{document}